\documentclass[times,twocolumn,final]{elsarticle}

\usepackage[table,xcdraw]{xcolor}

\usepackage{medima}
\usepackage{framed,multirow}
\usepackage{rotating}

\usepackage{zref-abspage,zref-user}
\usepackage{atbegshi}
\makeatletter
\newcommand{\rotatepageforlabelcw}[1]{%
  \zlabel{#1}%
  \AtBeginShipout{%
  \ifnum\c@page=\zref@extractdefault{#1}{abspage}{0}
    \pdfpageattr{/Rotate 90}
  \fi}
}
\newcommand{\rotatepageforlabelccw}[1]{%
  \zlabel{#1}%
  \AtBeginShipout{%
  \ifnum\c@page=\zref@extractdefault{#1}{abspage}{0}
    \pdfpageattr{/Rotate 270}
  \fi}
}
\makeatother

\usepackage{amssymb}
\usepackage{latexsym}
\usepackage{amsmath}
\usepackage{url}
\usepackage{xspace}

\usepackage{bm}

\usepackage{lineno,microtype,subcaption}
\usepackage{booktabs}
\usepackage{placeins}

\newcommand{\londonscgk}{London SC-GK\xspace}
\newcommand{\londonmcrc}{UK MC-RC\xspace}
\newcommand{\tilburgscgk}{Tilburg SC-GK\xspace}
\newcommand{\crossmoda}{crossMoDA\xspace}
\newcommand{\figref}[1]{\figurename~\ref{#1}}
\newcommand{\tabref}[1]{\tablename~\ref{#1}}

\definecolor{ne2e-c}{rgb}{0.122,0.467,0.706}
\definecolor{MAI-c}{rgb}{0.682,0.78,0.91}
\definecolor{LaTIM-c}{rgb}{1.0,0.498,0.055}
\definecolor{Super-Polymerization-c}{rgb}{1.0,0.733,0.471}
\definecolor{A*DA-c}{rgb}{0.173,0.627,0.173}
\definecolor{fgh-365-c}{rgb}{0.596,0.875,0.541}
\definecolor{SJTU-EIEE-2-426Lab-c}{rgb}{0.839,0.153,0.157}
\definecolor{MBZUAI-VS-c}{rgb}{1.0,0.596,0.588}
\definecolor{HUST-CBIB-c}{rgb}{0.58,0.404,0.741}
\definecolor{skjp-c}{rgb}{0.773,0.69,0.835}
\definecolor{gabybaldeon-c}{rgb}{0.549,0.337,0.294}
\definecolor{Of-Men-and-Rabbits-c}{rgb}{0.769,0.612,0.58}
\definecolor{vandy365-c}{rgb}{0.325,0.007,0.64}
\definecolor{superpoly-c}{rgb}{0.546,0.039,0.647}
\definecolor{ccc1018-c}{rgb}{0.723,0.196,0.539}
\definecolor{hustcbib-c}{rgb}{0.86,0.361,0.407}
\definecolor{auda-c}{rgb}{0.955,0.533,0.285}
\definecolor{mbzuai-c}{rgb}{0.994,0.741,0.166}
\definecolor{LightGray}{gray}{0.95}

\usepackage{enumitem}
\usepackage{pifont}
\newcommand{\xmark}{\ding{55}}%
\newcolumntype{P}[1]{>{\centering\arraybackslash}p{#1}}
\usepackage{adjustbox}

\renewcommand\cite{\citep}

\usepackage{hyperref}

\journal{Medical Image Analysis}

\begin{document}

\verso{Wijethilake, Dorent \textit{et~al.}}

\begin{frontmatter}

\title{\crossmoda Challenge: Evolution of Cross-Modality Domain Adaptation Techniques for Vestibular Schwannoma and Cochlea Segmentation from 2021 to 2023}%
% \tnotetext[tnote1]{This is an example for title footnote coding.}

\author[1]{Navodini \snm{Wijethilake}\texorpdfstring{\fnref{fn1}}{}}
\author[2]{Reuben \snm{Dorent}\texorpdfstring{\fnref{fn1}}{}}

\fntext[fn1]{Navodini Wijethilake and Reuben Dorent contributed equally to this work.}
%% Third author's email
% \ead{author3@author.com}
\author[1]{Marina \snm{Ivory}} %marina.ivory@kcl.ac.uk
\author[1]{Aaron \snm{Kujawa}} %aaron.kujawa@kcl.ac.uk
\author[3]{Stefan \snm{Cornelissen}}  %stefan.cornelissen@etz.nl
\author[3]{Patrick \snm{Langenhuizen}} %p.langenhuizen@etz.nl
\author[4]{Mohamed \snm{Okasha}} %mohamed.okasha@nhs.scot
\author[4]{Anna \snm{Oviedova}} %aoviedova@nhs.net
\author[5]{Hexin \snm{Dong}} %donghexin@pku.edu.cn
\author[7]{Bogyeong \snm{Kang}} % kangbk@korea.ac.kr
\author[8]{Guillaume \snm{Sallé}} %guillaume.salle@univ-brest.fr
\author[9,10]{Luyi \snm{Han}} %l.han@nki.nl
\author[11,12,13]{Ziyuan \snm{Zhao}} %Zhao_Ziyuan@i2r.a-star.edu.sg
\author[14]{Han \snm{Liu}} %han.liu@Vanderbilt.Edu
\author[14]{Yubo \snm{Fan}} %yubo.fan@vanderbilt.edu
\author[15]{Tao \snm{Yang}} %yangtao22@sjtu.edu.cn
\author[16]{Shahad \snm{Hardan}} %shahad.hardan@mbzuai.ac.ae
\author[16]{Hussain \snm{Alasmawi}} %hussain.alasmawi@mbzuai.ac.ae
\author[16]{Santosh \snm{Sanjeev}} %Santosh.Sanjeev@mbzuai.ac.ae
\author[17]{Yuzhou \snm{Zhuang}} %zhuang_yuzhou@hust.edu.cn
\author[19]{Satoshi \snm{Kondo}} %kondo@mmm.muroran-it.ac.jp
\author[21]{Maria \snm{Baldeon Calisto}} %mbaldeonc@usfq.edu.ec
\author[23]{Shaikh Muhammad Uzair \snm{Noman}} %U.Noman@campus.lmu.de
\author[24]{Cancan \snm{Chen}} %ccancan@infervision.com
\author[14]{Ipek \snm{Oguz}} %ipek.oguz@Vanderbilt.Edu
\author[24,25]{Rongguo \snm{Zhang}} %zrongguo@infervision.com
\author[23]{Mina \snm{Rezaei}} %Mina.Rezaei@stat.uni-muenchen.de
\author[22]{Susana K. \snm{Lai-Yuen}} %laiyuen@usf.edu
\author[20]{Satoshi \snm{Kasai}} 
\author[26]{Yunzhi \snm{Huang}} %yunzhi.huang.scu@gmail.com
\author[18]{Chih-Cheng \snm{Hung}} %chung1@kennesaw.edu
\author[16]{Mohammad \snm{Yaqub}} %mohammad.yaqub@mbzuai.ac.ae
\author[15]{Lisheng \snm{Wang}} %lswang@sjtu.edu.cn
\author[14]{Benoit M. \snm{Dawant}} %benoit.dawant@vanderbilt.edu
\author[13]{Cuntai \snm{Guan}} %ctguan@ntu.edu.sg
\author[9,10]{Ritse \snm{Mann}} %ritse.mann@radboudumc.nl
\author[8]{Vincent \snm{Jaouen}} %vincent.jaouen@imt-atlantique.fr
\author[7]{Tae-Eui \snm{Kam}} %kamte@korea.ac.kr
\author[5,6]{Li \snm{Zhang}} %zhangli_pku@pku.edu.cn
\author[1,4]{Jonathan \snm{Shapey}} %jonathan.shapey@kcl.ac.uk
\author[1]{Tom \snm{Vercauteren}} %tom.vercauteren@kcl.ac.uk
\footnotesize{
\address[1]{School of BMEIS, King's College London, London, United Kingdom}
\address[2]{Harvard University, USA}
\address[3]{Elisabeth-TweeSteden Hospital, Tilburg, Netherlands}
\address[4]{King's College Hospital, London, United Kingdom}
\address[5]{Center for Data Science, Peking University, Beijing, China}
\address[6]{Center for Data Science in Health and Medicine, Peking University, Beijing, China}
\address[7]{Department of Artificial Intelligence, Korea University, Seoul, Republic of Korea}
\address[8]{UMR 1101 Inserm LaTIM, Universit{\'e} de Bretagne Occidentale, IMT Atlantique, Brest, France}
\address[9]{Department of Radiology and Nuclear Medicine, Radboud University Medical Center, Geert Grooteplein 10, 6525 GA, Nijmegen, The Netherlands}
\address[10]{Department of Radiology, The Netherlands Cancer Institute, Plesmanlaan 121, 1066 CX, Amsterdam, The Netherlands}
\address[11]{Institute for Infocomm Research ($I^{2}$R), A*STAR, Singapore}
\address[12]{Artificial Intelligence, Analytics And Informatics ($AI^{3}$), A*STAR, Singapore}
\address[13]{Nanyang Technological University, Singapore}
\address[14]{Vanderbilt University, USA}
\address[15]{Department of Automation, Shanghai Jiao Tong University, Shanghai, China}
\address[16]{Mohamed bin Zayed University of Artificial Intelligence, Abu Dhabi, UAE}
\address[17]{School of Computer Science and Technology, Huazhong University of Science and Technology, Wuhan, China}
\address[18]{Center for Machine Vision and Security Research, Kennesaw State University, Marietta, MA 30060, USA}
\address[19]{Muroran Institute of Technology, Hokkaido, Japan}
\address[20]{Niigata University of Health and Welfare, Niigata, Japan}
\address[21]{Universidad San Francisco de Quito, Diego de Robles s/n y V{\'i}a Interoce{\'a}nica, Quito, Ecuador}
\address[22]{University of South Florida, Tampa, FL, USA}
\address[23]{Ludwig-Maximilians-Universit{\"a}t M{\"u}nchen, Germany}
\address[24]{Infervision Advanced Research Institute, Beijing, China}
\address[25]{Academy for Multidisciplinary Studies, Capital Normal University, Beijing, China}
\address[26]{School of Automation, Nanjing University of Information Science and Technology,
Nanjing 210044, China}
}
%\received{1 May 2013}
%\finalform{10 May 2013}
%\accepted{13 May 2013}
%\availableonline{15 May 2013}
%\communicated{S. Sarkar}

\begin{abstract}
%%%
The cross-Modality Domain Adaptation (\crossmoda) challenge series, initiated in 2021 in conjunction with the International Conference on Medical Image Computing and Computer Assisted Intervention (MICCAI), focuses on unsupervised cross-modality segmentation, learning from contrast-enhanced T1 (ceT1) and transferring to T2 MRI.
The task is an extreme example of domain shift chosen to serve as a meaningful and illustrative benchmark.  From a clinical application perspective, it aims to automate Vestibular Schwannoma (VS) and cochlea segmentation on T2 scans for more cost-effective VS management.
The challenge has evolved across three editions to address increasingly complex clinical scenarios: beginning with single-institution controlled data for binary segmentation (tumour and cochlea) in 2021, extending to multi-institution data with Koos grade classification in 2022, and culminating in heterogeneous routine surveillance data with intra- and extra-meatal tumour sub-segmentation in 2023. 
In this work, we report the findings of the 2022 and 2023 \crossmoda editions alongside a retrospective analysis of challenge progression. Successive editions demonstrate a reduction in segmentation outliers despite concurrently increasing dataset heterogeneity, suggesting that expanded training diversity improves model robustness. Notably, the 2023 winning approach generalised effectively to prior homogeneous test sets, evidencing the benefit of heterogeneous training data. 
However, cochlear Dice scores declined in 2023. This decrease may reflect the combined effects of increased dataset heterogeneity, resolution variability, and the added optimisation complexity of a three-class segmentation task, although the precise cause remains uncertain.
Progress can and should still be made on the specific task of VS segmentation to reach clinical acceptability but as the performance of leading competitors starts to plateau, a more challenging cross-modal learning task may be beneficial to serve as a benchmarking tool in the future.
%%%%
\end{abstract}
%\begin{keyword}
%\end{keyword}

\end{frontmatter}

%\linenumbers

%% main text
\section{Introduction} \label{sec1:intro}
%problem 
Machine learning has made significant strides in medical image analysis. Model development is typically based on the assumption that training and testing datasets share similar distributions. However, practical clinical settings often involve variations in acquisition protocols, leading to \emph{domain shift} and compromising model performance. 
This issue is particularly pronounced in deep learning models like Convolutional Neural Networks, which are highly sensitive to distribution shifts, especially when trained on limited data, as is often the case in medical imaging. To address the robustness challenges raised by domain shifts, domain adaptation (DA) has emerged as a promising approach. 
DA techniques aim to improve model robustness by aligning distributions between training and target domains. A large variety of methods has been proposed for medical image tasks, including supervised, semi-supervised, weakly supervised, and unsupervised approaches. Among these, unsupervised methods are particularly appealing, as they do not require additional annotations from medical experts. A key challenge is to ensure that these unsupervised techniques generalise effectively to heterogeneous, real-world clinical datasets. Therefore, validating DA methods using public benchmarks on diverse routine clinical data is crucial to assess their reliability and practical applicability.

In medical imaging, domain shifts across medical sites or scanners have been frequently analysed, whereas cross-modality DA remains relatively unexplored.
As an extreme case of domain shift, designing a public benchmark around cross-modality DA serves as a meaningful and illustrative task.
The choice of technical challenge should nonetheless not be made at the expense of clinical relevance.
For aggressive tumours, both ceT1 and T2 modalities are typically acquired, however, in the routine clinical management of benign tumours, the choice of modality is site dependent, with either ceT1 or T2 being prioritised. 
As a relevant example of where cross-modality DA can have a positive clinical impact,
this challenge addresses imaging of a specific benign brain tumour: VS.
VS originates from the Schwann cells of the vestibular nerve sheath. 
Currently, ceT1 MR imaging is widely used for diagnosing and monitoring VS. However, due to safety concerns associated with gadolinium-based contrast agents, there is growing interest in using non-contrast imaging sequences such as T2 imaging \cite{connor2021imaging}. In addition to improving patient safety, T2 imaging is ten times more cost-efficient than ceT1 imaging \cite{connor2021imaging}.
%
%importance of segmentation of VS
VS has an estimated incidence of 1 in 1000 \cite{evans2005incidence} and for these smaller tumours, expectant management with observation using MR imaging is often advised. 
The choice of treatment is typically determined by tumour size and its impact on adjacent brain structures, such as the brainstem and cranial nerves. The Koos grading scale is a classification system for VS that reflects these key characteristics that influence treatment decisions \citep{koos1998neurotopographic}.
If the tumour demonstrates growth, management options include conventional open surgery or stereotactic radiosurgery. Accurate segmentation of VS and surrounding critical structures such as the cochlea is crucial for effective treatment planning and follow-up \cite{shapey2018standardised}. Traditionally, tumour growth is estimated by measuring its maximal linear dimension \citep{kanzaki2003new}. The Consensus Meeting on Systems for Reporting Results in Acoustic Neuroma (also known as VS) recommends clearly distinguishing between the intra- and extra-meatal portions of the tumour, where the tumour is separated respectively into inner ear and outer ear regions, along the posterior wall of the petrous pyramid \cite{kanzaki2003new}. Then the largest extra-meatal diameter used for reporting tumour size in current clinical settings \cite{kanzaki2003new}. This highlights the need for segmenting intra- and extra-meatal regions within the VS tumour, to establish a reliable routine for reporting. Moreover, recent studies suggest that volumetric measurements provide a more accurate and sensitive assessment of VS size, particularly for detecting subtle growth \cite{fink2022comparing}. Therefore, accurate VS segmentation, along with a clear distinction between intra- and extra-meatal regions, is essential for extracting standardised imaging biomarkers and for accurate growth assessment.

In light of these challenges in cross-modality DA and need for precise tumour volumetric measurements, we proposed a cross-modality benchmark that transitions from ceT1 to T2 imaging, aiming to automate the segmentation of both VS and cochleas on T2 scans. We created the \crossmoda challenge in 2021 in conjunction with the 24th International Conference on Medical Image Computing and Computer-Assisted Intervention (MICCAI 2021). The primary objective of this challenge was to benchmark both new and existing unsupervised DA techniques specifically designed for medical image segmentation.

By leveraging data from 379 patients from a single institute, the \crossmoda 2021 challenge aimed to develop robust segmentation methods for two critical brain structures: the VS and the cochleas. This pioneering effort represents the first large-scale, multi-class benchmark focused on unsupervised cross-modality DA in medical imaging. 

The 2022 edition expanded the scope of the challenge significantly. Compared to the previous \crossmoda 2021 challenge, which utilised single-institution data and featured a single segmentation task, it included multi-institutional data acquired under controlled conditions for radiosurgery planning and introduced an additional classification task for grading VS according to the Koos grade~\citep{koos1998neurotopographic}.

The 2023 edition further extended the segmentation task by incorporating multi-institutional, heterogeneous data obtained for routine surveillance purposes, introducing a significant resolution shift compared to datasets from previous editions. Additionally, this edition introduced a sub-segmentation task for the tumour, distinguishing between intra- and extra-meatal components, thereby creating a three-class problem. 
%These enhancements in the \crossmoda challenge aimed to advance the development of effective techniques capable of addressing the inherent challenges posed by domain shifts in medical image analysis.

Over these three editions, we progressively increased the challenge complexity, evolving from a one-to-one to a one-to-many domain shift scenario. The \crossmoda challenge has enabled comparing and validating a broad range of unsupervised DA methods in practical, real-world settings where target data is heterogeneous. While a detailed review of DA methodologies is beyond the scope of this paper, we direct interested readers to \citep{guan2021domain} for a comprehensive overview. Our previous \crossmoda 2021 paper \citep{dorent2023crossmoda} extensively discussed the benchmark datasets used to assess DA techniques for unsupervised medical image segmentation. 
Cross-modality DA in medical imaging has been approached through several complementary methodological families. A widely used strategy is image-level adaptation, where source-domain images are translated into the appearance of the target modality using unpaired image-to-image translation methods such as CycleGAN~\citep{zhu2017unpaired} and its variants. These approaches aim to reduce the visual domain gap while preserving anatomical content, and have been widely adopted in cross-modality segmentation tasks. 
A second family of approaches focuses on feature-level adaptation, where adversarial learning, contrastive learning, or distribution alignment losses are used to encourage modality-invariant representations within the segmentation network~\citep{kamnitsas2017unsupervised}. 
More recent methods combine image- and feature-level adaptation with self-training, where pseudo-labels generated on unlabelled target-domain images are iteratively refined to improve target-domain performance, as illustrated by the \crossmoda 2021 top-performing approach~\citep{shin2022cosmos}. 
Consistency regularisation and mean-teacher frameworks have also been used to stabilise pseudo-label learning under image perturbations or multi-scale predictions~\citep{perone2019unsupervised}. 
Finally, segmentation-aware synthesis has emerged as a particularly relevant direction for medical imaging, as it explicitly constrains image translation to preserve small or clinically important structures~\citep{chen2020unsupervised}. 
The methods submitted to the \crossmoda challenge reflect many of these developments, while providing a controlled benchmark to compare their robustness under progressively more heterogeneous and clinically realistic cross-modality settings.
%

% \section{Related Work}

% The participating teams employed a wide range of DA methods, illustrating the breadth of approaches in this field. While a detailed review of DA methodologies is beyond the scope of this paper, we direct interested readers to \citep{guan2021domain} for a comprehensive overview.
% We also refer to our previous \crossmoda 2021 paper \citep{dorent2023crossmoda}, for an extensive discussion of the benchmark datasets used to assess DA techniques for unsupervised medical image segmentation. 

%discussing the related work on DA on heterogenous data. 

% \textcolor{red}{TODO: Add a paragraph on the contributions of this paper. The abstract is a good starting point for a structure. Also clarify the paper organisation.}
In this work, we summarise the approaches used in the 2022 and 2023 \crossmoda editions and conduct a retrospective analysis of the challenge's evolution, highlighting the contributions of each edition. First, we detail the challenge's expansion over three years and the datasets introduced in each edition. We also present the metrics used in the challenge and the approach followed for ranking the teams. Then, we provide a detailed overview of the techniques used by participants in each edition. 
This is followed by a detailed analysis of segmentation performance, including stratification for each structure, contribution of different components to the overall ranking, and ranking stability. Finally, we discuss the improvements observed in DA techniques over time and retrospectively assess the winning models' performance across editions. We conclude by discussing the limitations and the possible future expansions of the \crossmoda challenge. 

\section{Challenge description}
\subsection{Overview}
The \crossmoda challenge series aims to benchmark both new and existing techniques for unsupervised cross-modality DA in medical image segmentation (2021-2023) and classification (2022). Initially, the challenge focused on segmenting two critical brain structures involved in the monitoring and treatment planning of VS: the tumour and the cochleas. Participants were required to develop algorithms capable of performing inference on high-resolution T2 scans. For this challenge, the training set comprised high-resolution T2 scans \emph{without} manual annotations, while manual annotations were provided for an unpaired training set of ceT1 scans.  Thus, participants needed to accomplish unsupervised cross-modality DA from ceT1 (source) scans to T2 (target) scans. However, the \crossmoda inaugural challenge in 2021 was limited to single institutional data acquired under controlled conditions for radiosurgery planning, with high-resolution T2 scans in the target set.

Over time, the objectives of the \crossmoda challenge have evolved significantly, aiming to enhance the generalisation and robustness of the models. Compared to the initial edition, which utilised single-institutional data and focused on a two-class segmentation task (tumour and cochlea), the 2022 challenge expanded the dataset to include multi-institutional data and introduced an additional classification task for grading VS according to the Koos grade. This marked an important enhancement allowing to validate the adaptability of techniques across different institutions and imaging conditions. However, the datasets introduced in 2021 and 2022 consisted of high-resolution scans as both were acquired for treatment planning purposes under controlled conditions.

The 2023 edition further extended the segmentation task by incorporating a multi-institutional and highly heterogeneous routine surveillance dataset. 
The colossal increase in data heterogeneity illustrated in \figref{fig:dataset_heterogeneity} not only made the task more challenging but it also helped demonstrate the real-world importance of cross-modality DA.
Indeed, unlike for treatment planning, in routine clinical scenarios, acquiring both ceT1 and T2 scans is unconventional. 
This edition also introduced a sub-segmentation task for the tumour, distinguishing between intra- and extra-meatal components, thereby creating a three-class problem. This progression from controlled radiosurgery planning data to heterogeneous routine surveillance data reflects an ongoing effort to address more complex and clinically relevant scenarios with impact beyond VS. 

\subsection{Data description}

\subsubsection{Data overview}
The dataset utilised in this study is derived from three distinct sources as detailed in \tabref{tab1}:
\begin{itemize}
    \item \londonscgk \cite{shapey2021segmentation}: Treatment planning MRI from Queen Square Radiosurgery Centre, London, United Kingdom 
    \item \tilburgscgk: Treatment planning MRI from TweeSteden Hospital, Tilburg, Netherlands 
    \item \londonmcrc \citep{aaronnmm}: A compilation of routine imaging data from ten medical sites across the United Kingdom
    %, with imaging dates spanning from February 2006 to September 2019 
\end{itemize} 
The \crossmoda 2021 challenge started with the \londonscgk dataset only. The \tilburgscgk data was introduced for the 2022 edition, making the combined dataset multi-institutional.
However, the \londonscgk and \tilburgscgk datasets were acquired for radiosurgery treatment planning, therefore acquired under controlled conditions.  
For the 2023 edition, we introduced the \londonmcrc dataset, where the scans were acquired for routine surveillance purposes without any predefined controlled conditions, making the combined dataset much more heterogeneous. 

For 2021 and 2022 segmentation tasks, we included patients who had undergone surgery. For the classification task in 2022, we excluded cases who had
previously undergone operative surgical treatment, as Koos grade is most useful used pre-operatively.
%hese diverse sources provide a comprehensive and varied dataset, critical for robust domain adaptation in medical image analysis. 
Further, for the 2023 challenge, we excluded the postoperative scans, as surgical procedures could affect the anatomical distinction between intra- and extra-meatal regions. 

\begin{figure*}[htbp!]
\begin{center}
\includegraphics[width=0.8\textwidth]{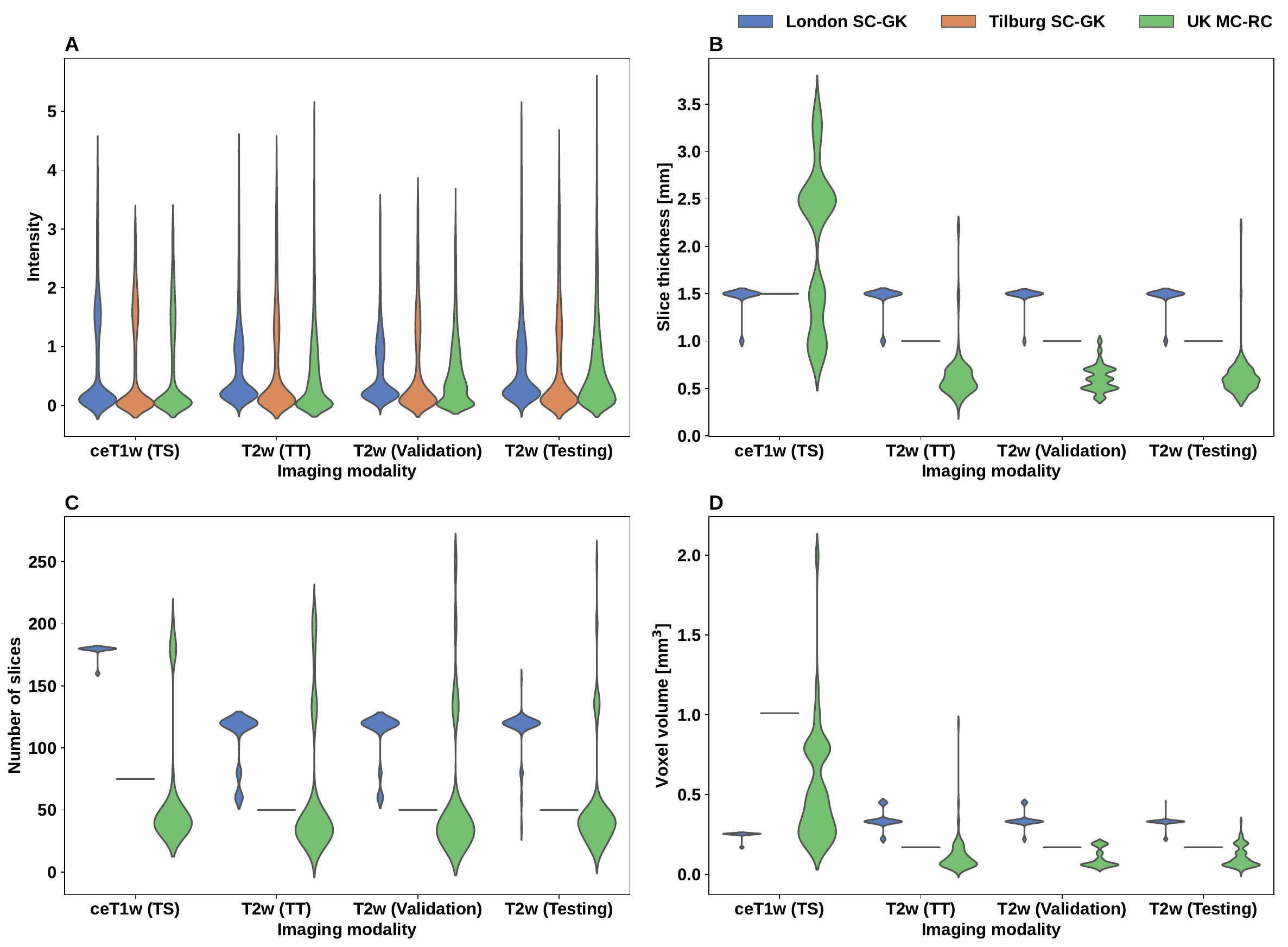}
\end{center}
\caption{Distribution of (A) Intensity (B) Voxel volume [$mm^3$] (C) Number of slices (D) Slice thickness [$mm$], within 3 datasets used in the \crossmoda challenge. \londonscgk, \tilburgscgk and \londonmcrc datasets were introduced in respectively 2021, 2022, and 2023 editions of \crossmoda challenge.}\label{fig:dataset_heterogeneity}
\end{figure*}

\begin{table*}[thbp!]
\caption{\label{tab1}Distribution of data within the training, validation and testing splits from 3 data sources used in the three editions of the \crossmoda challenge.}
\centering
\resizebox{0.75\textwidth}{!}{
\begin{tabular}{ccP{1.2cm}P{1.2cm}P{2.5cm}P{2.5cm}c}
\toprule
\multirow{2}{*}{\textbf{Dataset}}  & \multirow{2}{*}{\textbf{Edition}} &  \multicolumn{2}{c}{\textbf{\londonmcrc}} & \multicolumn{1}{c}{\textbf{\londonscgk}} & \multicolumn{1}{c}{\textbf{\tilburgscgk}} & \multirow{2}{*}{$\mathbf{N_{total}}$} \\ 
\cmidrule(r){3-4} \cmidrule(lr){5-6} 
  & &$N_{patients}$ & $N_{scans}$ & $N_{scans}(=N_{patients})$ & $N_{scans}(=N_{patients})$ &  \\ 
\toprule
% \multicolumn{6}{l}{\textbf{Training Source Data }}		\\ %\midrule
% % \multicolumn{6}{l}{\tab \textbf{Source Domain	- ceT1}}	\\ %\midrule
 \rowcolor{LightGray} \cellcolor{white} & \crossmoda 2021 & - &- &	105&	- &	105
\\% \midrule
\textbf{Training Source Data } & \crossmoda 2022 & - &- &	105 &	105 &	210
\\% \midrule
\rowcolor{LightGray} \cellcolor{white} & \crossmoda 2023  & 38	&	43	&	79	&	105	&	227 \\ \midrule
 %\midrule	
\rowcolor{LightGray} \cellcolor{white} & \crossmoda 2021	&	-	&	-		& 105	&	-	&	105 \\ 
\textbf{Training Target Data } & \crossmoda 2022	&	-	&	-		& 105	&	105 &	210 \\
\rowcolor{LightGray} \cellcolor{white} & \crossmoda 2023	&	47	&	105	&	85	&	105	&	295 \\ \midrule
 \rowcolor{LightGray} \cellcolor{white} & \crossmoda 2021	&	-	&	-		& 32	&	-	&	32 \\ 
\textbf{Validation Target Data } & \crossmoda 2022	&	-	&	-		& 32	&	32 &	64 \\
 \rowcolor{LightGray} \cellcolor{white} &
\crossmoda 2023	& 15	&	32		& 32	&	32	&	96 \\ \midrule
 \rowcolor{LightGray} \cellcolor{white} & \crossmoda 2021	&	-	&	-		& 137	&	-	&	137 \\ 
\textbf{Testing Target Data} & \crossmoda 2022	&	-	&	-		& 137	&	134 &	271 \\
 \rowcolor{LightGray} \cellcolor{white} &  \crossmoda 2023	& 45	&	94	&	113	&	134	&	341\\ \midrule
 \rowcolor{LightGray} \cellcolor{white} & \crossmoda 2021	&	-	&	-		& 379	&	-	&	379 \\ 
\textbf{Total Data} & \crossmoda 2022	&- &- &	379&	376 &	755\\
\rowcolor{LightGray} \cellcolor{white} & \crossmoda 2023	&145	&	274	&	309	&	376	&	959\\ 
\bottomrule
\end{tabular}}
\end{table*}

\subsubsection{Image acquisition}

\paragraph{\londonscgk} 
The contrast-enhanced T1-weighted imaging for the London SC-GK dataset was performed using an MP-RAGE sequence. This sequence had an in-plane resolution of 0.47×0.47 mm, an in-plane matrix of 512×512, and a slice thickness ranging from 1.0 to 1.5 mm (TR=1900 ms, TE=2.97 ms, TI=1100 ms). High-resolution T2-weighted imaging was conducted with either a 3D CISS or FIESTA sequence, offering an in-plane resolution of 0.5×0.5 mm, an in-plane matrix of 384×384 or 448×448, and a slice thickness between 1.0 and 1.5 mm (TR=9.4 ms, TE=4.23 ms). The scans were acquired using a 32-channel Siemens Avanto 1.5T scanner using a Siemens
single-channel head coil.

\paragraph{\tilburgscgk} 
For the Tilburg SC-GK dataset, the contrast-enhanced T1-weighted imaging was carried out using a 3D-FFE sequence. This sequence had an in-plane resolution of 0.8×0.8 mm, an in-plane matrix of 256×256, and a slice thickness of 1.5 mm (TR=25 ms, TE=1.82 ms). The high-resolution T2-weighted imaging was performed using a 3D-TSE sequence, with an in-plane resolution of 0.4×0.4 mm, an in-plane matrix of 512×512, and a slice thickness of 1.0 mm (TR=2700 ms, TE=160 ms, ETL=50). The scans were acquired with a Philips Ingenia 1.5T scanner using a Philips quadrature head coil. 

\paragraph{\londonmcrc}
The UK MC-RC dataset, stemming from 10 different sites with imaging dates spanning February 2006 to September 2019, exhibited significant variation in slice thickness, voxel volume, and intensities across all contrast-enhanced T1-weighted and T2-weighted imaging. This variability is illustrated in \figref{fig:dataset_heterogeneity}, highlighting the heterogeneity of the imaging parameters used across different institutions and timeframes. The scans were acquired using a range of MRI scans including, SIEMENS, Philips, General Electrics, Hitachi MRI scanners with the magnetic field strengths 1.0T/1.5T/3.0T.

\paragraph{Ethics statement} This study was approved by the NHS Health Research Authority and Research Ethics Committee (18/LO/0532). Because the study was a retrospective one and the MR images were de-identified before analysis, no informed consent was required from the study participants.

\subsection{Reference segmentation protocol}

\paragraph{\londonscgk and \tilburgscgk}
All imaging datasets were manually segmented following the same annotation protocol. 
The tumour volume (VS) was manually segmented by the treating neurosurgeon and physicist using both the ceT1 and T2 images.
All VS segmentations were performed as part of standard clinical practice using the Leksell GammaPlan software that employs an in-plane semi-automated segmentation method. Using this software, delineation was performed on sequential 2D axial slices to produce 3D models for each structure. 

The adjacent cochlea (hearing organ) is the main organ at risk during VS radiosurgery. In the 2021 \crossmoda dataset, patients have a single sporadic VS. Consequently, only the cochlea closest to the tumour was initially segmented by the treating neurosurgeon and physicist as part of standard clinical practice. 
Preliminary results using a fully-supervised approach \citep{isensee2021nnu} showed that considering the remaining cochlea as part of the background led to poor performance for cochlea segmentation. 
Given that tackling this challenging issue is beyond the scope of the challenge, both cochleas were manually segmented by radiology fellows with over 3 years of clinical experience in general radiology using the ITK-SNAP software~\citep{py06nimg}. 
T2 images were used as reference for cochlea segmentation. The basal turn with osseous spiral lamina was included in the annotation of every cochlea to keep manual labels consistent. In addition, modiolus, a small low-intensity area (on T2) within the centre of the cochlea, was included in the segmentation as well. 

For the 2022 edition, automated GIF~\citep{cardoso2015geodesic} parcellation masks were provided for the training source dataset, as they have been shown to be of interest for performing automated Koos classification~\cite{kujawa2022automated}. Participants were allowed to use these masks for both tasks. 

For the 2023 edition, a trained radiologist sub-segmented the VS tumour regions in \londonscgk and \tilburgscgk datasets into intra- and extra-meatal regions using the ITK-SNAP tool \cite{py06nimg}.

\paragraph{\londonmcrc} 
This dataset
%The original dataset, available on TCIA (The Cancer Imaging Archive)\footnote{\url{https://doi.org/10.7937/HRZH-2N82}},
%\url{https://www.cancerimagingarchive.net/collection/vestibular-schwannoma-mc-rc/}, 
comprises of MRI scans from 165 patients with unilateral VS \citep{aaronnmm}. Whole tumour annotation was conducted iteratively by a specialist MRI labeling company, Neuromorphometrics (Somerville, Massachusetts, USA). This annotation was then reviewed and validated by a panel of clinical experts, including a consultant neuroradiologist and a consultant neurosurgeon with the  inter- and intrarater variability analysis, with an analysis of Inter- and intra-observer reliability. A detailed explanation of this process may be found in previous work \cite{aaronnmm}.

The ground truth for the cochlea and VS in the UK MC-RC dataset (training-source) was independently annotated from scratch on ceT1 and T2 images, without cross-referencing between modalities, using ITK-SNAP software \citep{py06nimg}. Unlike in the other dataset annotations, only the ceT1 with clearly visible cochlea regions were included. However, due to the limited visibility of the cochlea region on the ceT1 images, minor annotation inaccuracies may be present. For the cochlea segmentation on the T2 MR images from the UK MC-RC dataset, the basal turn with osseous spiral lamina was included in the annotation of every cochlea to keep manual labels consistent. In addition, modiolus, a small low-intensity area (on T2) in the centre of the cochlea, was also included in the segmentation. The annotations were performed by a medically trained researcher with over three years of clinical experience in general radiology.

Subsequently, intra-/extra-meatal segmentation (split segmentation) within the tumour region was performed by an expert neurosurgeon using MRtrix3~\cite{tournier2019mrtrix3}, which was later quality controlled by a trained radiologist using ITK-SNAP tool~\cite{py06nimg}. For the VS on the T2 MR images from the UK MC-RC dataset, capping and intratumoural cysts were included in the segmentation. Other anatomical structures, such as blood vessels and other cranial nerves, were not labelled if found adjacent to the VS or within the tumour volume.

\subsection{Reference Koos grading protocol}

\begin{figure*}[ht!]
    \centering
    \begin{subfigure}{0.48\textwidth}
        \centering
        \includegraphics[width=\linewidth]{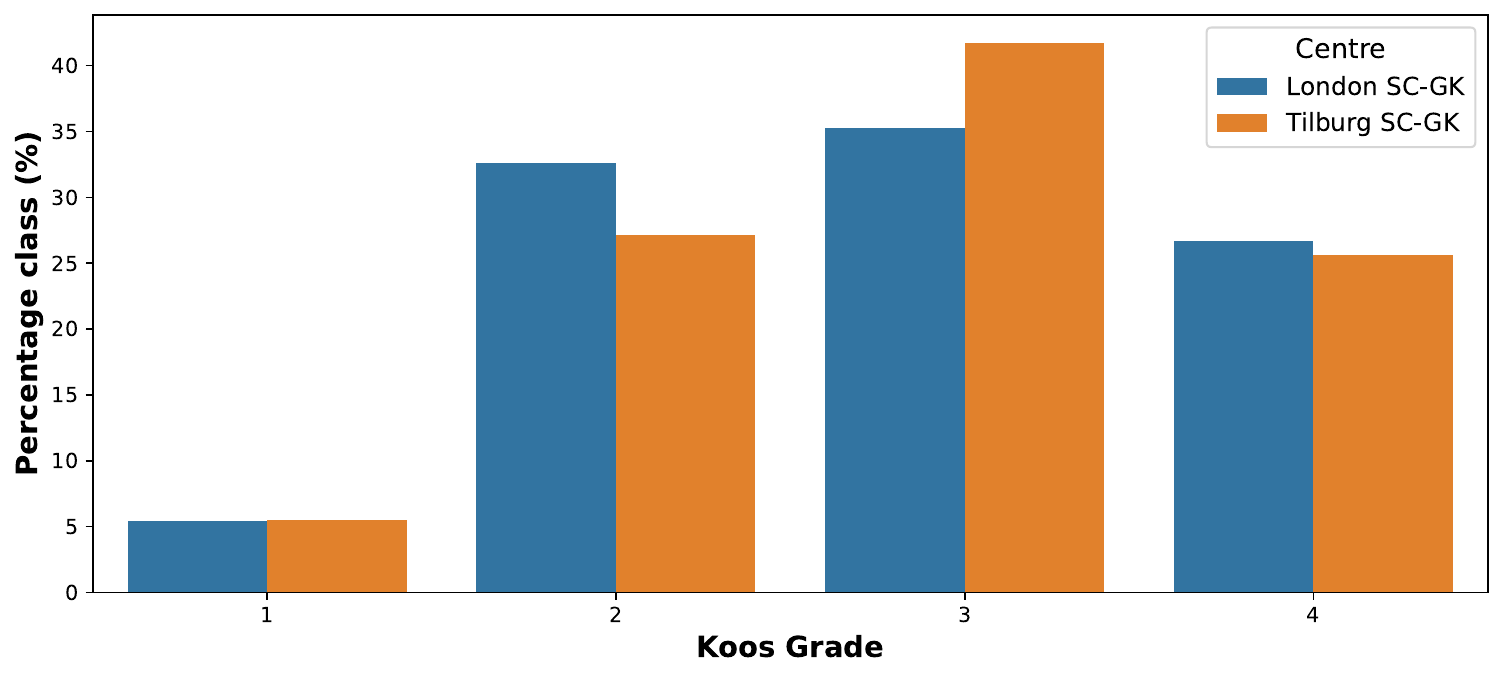}
        \caption{Distribution per clinical centre}
        \label{fig:distrib_koos_centre}
    \end{subfigure}
    \hfill
    \begin{subfigure}{0.48\textwidth}
        \centering
        \includegraphics[width=\linewidth]{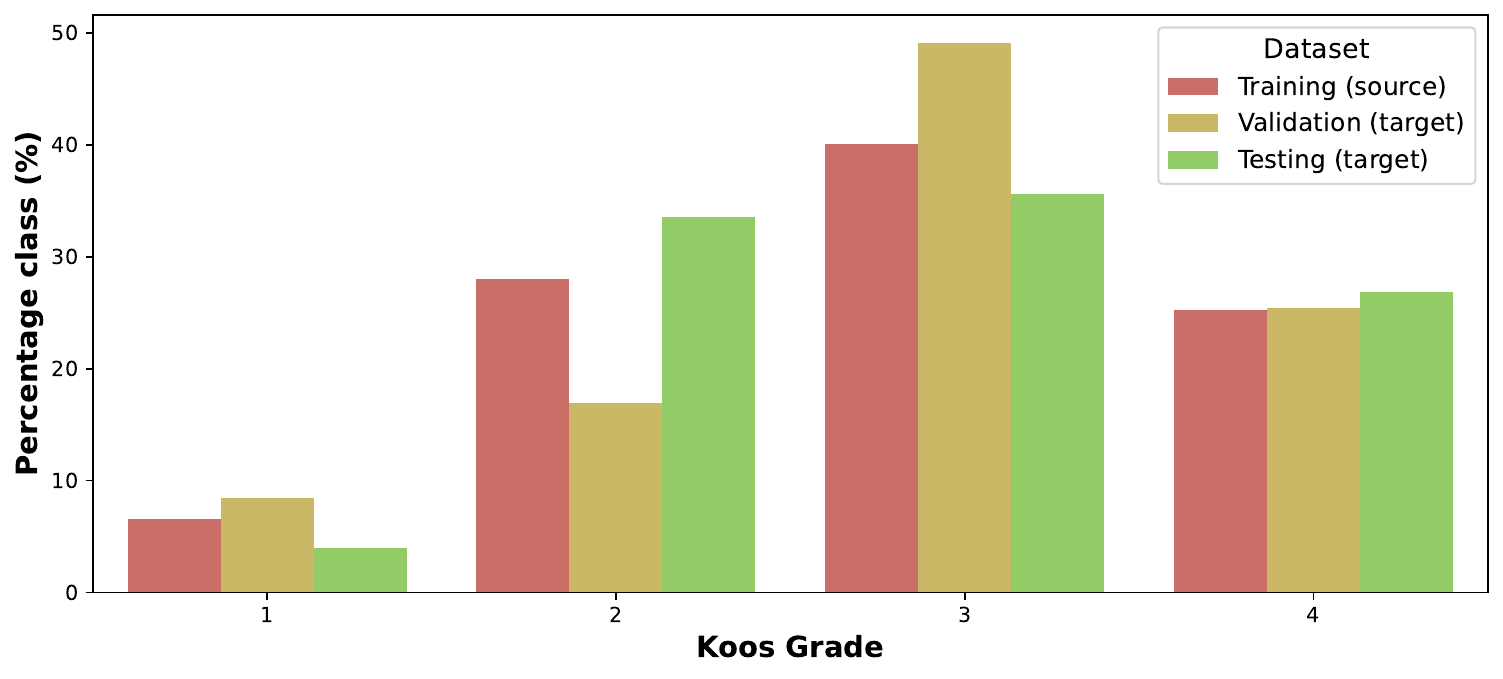}
        \caption{Distribution within the training/validation/testing subsets}
        \label{fig:distrib_koos_dataset}
    \end{subfigure}
    \caption{Stratification of Koos grade distributions in the \crossmoda 2022 dataset.\label{fig:distrib_koos}}
    \label{fig:main}
\end{figure*}

For the 2022 edition, two neurosurgeons with 5–10 years of experience independently assigned Koos grades to all pre-operative cases in the source training, target validation, and target test datasets from London SC-GK (N=221) and Tilburg SC-GK (N=273) in the \crossmoda 2022 dataset. Both contrast-enhanced T1-weighted and high-resolution T2-weighted MRI scans were provided for grading. Before annotation, each expert received a briefing along with image examples and a detailed description of the Koos grading system, as outlined by \citeauthor{erickson2019koos}. If the two annotators assigned the same grade, it was used as the ground truth. In cases of disagreement, a consensus grade was determined by a secondary team comprising a consultant neuroradiologist and a consultant neurosurgeon. These labels were also utilised in \citep{kujawa2022automated}.

In \figref{fig:distrib_koos_centre}, we present the distribution of Koos grades across the two centers. The distributions appear similar, with the lowest Koos grade (1) being underrepresented, as these cases are less likely to undergo stereotactic radiosurgery. For this reason, the classification task is imbalanced. In \figref{fig:distrib_koos_dataset}, we further compare the distribution of Koos grades across the three datasets: source training, target validation, and target testing. The distributions remain comparable across datasets.

\subsubsection{Data curation}

The three datasets used in this challenge were fully de-identified by removing the patient-identifiable information from the image headers and by defacing them. Defacing protocol of the \londonscgk can be found in \cite{shapey2021segmentation, milchenko2013obscuring}. The protocol followed for defacing \londonmcrc dataset is available at: \url{https://github.com/aaronkujawa/defacing_pipeline}.  All data were visually inspected before release.

The images and annotations were distributed in the NIFTI format (nii.gz). The 2022 edition datasets were released on zenodo\footnote{\url{https://zenodo.org/records/6504722}}. The datasets used in the 2023 edition were available on the Synapse \crossmoda challenge space. 

The training and validation data are distributed under the CC BY-NC-SA 4.0 (Attribution-NonCommercial-ShareAlike) license. 

Patient-level demographic metadata (e.g., age, sex, and ethnicity) were not released as part of the challenge datasets because they were either not consistently available across all sources or were excluded to minimise re-identification risk. Consequently, the challenge focuses on imaging rather than demographic heterogeneity. We acknowledge that the absence of demographic metadata limits subgroup and bias analyses and should be addressed in future benchmark datasets where such information can be shared safely.

\subsection{Challenge setup} 
For the 2021 and 2022 editions of the \crossmoda challenge Grand Challenge\footnote{\url{https://grand-challenge.org/}} platform was utilised and the 2023 edition was hosted on Synapse\footnote{\url{https://www.synapse.org/}}. Both platforms are well-established for managing automated validation leaderboards, but Synapse additionally allowed us to host our data on their platform. For the validation phase, we utilised the Synapse Workflow Orchestrator\footnote{\url{https://github.csupervisionom/Sage-Bionetworks/SynapseWorkflowOrchestrator}}, adhering to the data-to-model challenge workflow.

External data and pre-trained models were not allowed, but limited use of a generic brain atlas (e.g., for MNI normalisation) was permitted with clear justification. Members of the organising institutes could participate in the challenge but were not eligible for awards and were not listed on the leaderboard. The winning teams of each edition received an award. Rankings of all participating teams were publicly available on the official challenge website\footnote{\url{https://crossmoda-challenge.ml/}}, and the top 10 teams were invited to present their methods at the MICCAI BrainLes workshop. The first (co-)author(s) and senior author of the submitted short paper qualified as authors for the joint publication. Submission instructions were provided on Synapse\footnote{\url{https://www.synapse.org/Synapse:syn51236108}}.

To mitigate the risk that participants select their model hyper-parameters in a supervised manner, i.e. by computing the prediction accuracy, only one submission per day was allowed on the validation leaderboard. The validation phase of the \crossmoda 2022 was held between the 12th May 2022 and the 10th August 2022 and the evaluation phase was held between 5th August 2022 and 15th August 2022. The validation phase of the \crossmoda 2023 was held between the 27th April 2023 and the 9th of July 2023 and the evaluation phase was held between 5th July 2023 and 20th July 2023.

To evaluate the accuracy of predictions made on local machines, predictions on the validation set were computed and compared with those obtained from participants' machines. participants were required to include predictions for all cases in the submitted zip file. Submissions with incomplete predictions were not evaluated and were not counted as valid validation submissions. The evaluation code is available on github \footnote{\url{https://github.com/ReubenDo/crossmoda2022}}. 

Adhering to best practice guidelines for challenge organisation~\cite{maier2020bias}, the test set was kept private to mitigate the risk of cheating. Participants were required to containerise their methods using Docker, following the provided guidelines. Each team was permitted a single submission, and the Docker containers were executed on a cluster for evaluation. As a result, no missing results were expected for the test set. All participant containers successfully passed the quality control test. All participating teams were encouraged to provide an open-access to their training and testing code. 

\section{Metrics and evaluation}
\subsection{Choice of segmentation metrics} %same as reubens paper
The primary goal of the algorithms is to optimise the accuracy of the predictions. To avoid the limitations of relying on a single metric for segmentation assessment, we employed two metrics: the Dice Similarity Coefficient (DSC) and the Average Symmetric Surface Distance (ASSD), following the recent consensus recommendations on the selection of metrics for biomedical image analysis tasks \citep{maier2022metrics}. These metrics are also widely used in prior challenges due to their simplicity, rank stability, and effectiveness in evaluating segmentation accuracy.

The DSC measures the similarity between the predicted binary segmentation mask \(S_k\) and the manual segmentation \(G_k\) for a region \(k\), where \(k\) can be VS (intra-/extra-meatal/combined) or cochleas. 
It is calculated by normalising the size of their intersection over the average of their sizes:
\begin{equation}
\text{DSC}(S_k, G_k) = \frac{2 \sum_i S_{k,i} G_{k,i}}{\sum_i S_{k,i} + \sum_i G_{k,i}} \ .
\end{equation}

The ASSD evaluates the average Euclidean distance between the boundary of the predicted segmentation mask \(B_{S_k}\) and the boundary of the manual segmentation \(B_{G_k}\). It is computed as follows:
\begin{equation}
\text{ASSD}(S_k, G_k) = \frac{\sum_{s_i \in B_{S_k}} d(s_i, B_{G_k}) + \sum_{s_i \in B_{G_k}} d(s_i, B_{S_k})}{|B_{S_k}| + |B_{G_k}|} \ ,
\end{equation}
where \(d\) represents the Euclidean distance between points. If the predictions only contain background (i.e., \(S_{k,i} = 0\) for all voxels \(i\)), the ASSD is assigned the maximal distance between voxels in the test set (350mm).

\paragraph{\crossmoda 2023 edition} 
In the \crossmoda 2023 edition, as mentioned earlier, the challenge was extended to a three-class segmentation task, where the tumour was divided into intra- and extra-meatal regions in addition to the cochlea region. Consequently, DSC was measured separately for the intra- and extra-meatal regions and for cochlea. Additionally, for additional insight but not for ranking purposes, the overall DSC of VS was computed by combining the segmentation maps of both intra- and extra-meatal regions. 
%Similarly, for ranking, we used the combined segmentation maps of the intra- and extra-meatal regions (i.e., the VS tumour region) following the 2021 and 2022 ranking with 2 classes, instead of three.

In addition to the DSC, the ASSD metric was calculated for the 3 regions (intra- and extra-meatal regions and cochlea).
To get additional insight (rather than for ranking purposes), the ASSD was also computed for the VS region combining the intra- and extra-meatal regions, 
and specifically along the boundary between the intra- and extra-meatal regions. 
This focused evaluation aimed to accurately assess the precision of the intra-/extra-meatal boundary separation along the posterior wall of the petrous pyramid. 

\subsection{Choice of classification metric}
To evaluate the performance of different methods in the Koos classification task of the \crossmoda 2022 challenge, we employed the weighted macro-averaged mean absolute error (MA-MAE) metric~\citep{baccianella2009evaluation}. MA-MAE is particularly well-suited for imbalanced ordinal classification problems, such as Koos grading. It measures the absolute difference between the true $\bm{l}$ and predicted $\bm{\hat{l}}$ labels, ensuring that larger discrepancies between predicted and ground truth grades contribute to a higher classification error. Additionally, it accounts for class imbalance by performing a weighted average, where classes with fewer samples receive higher weights.
The weighted macro-averaged mean absolute error is defined as:
\begin{equation}
\text{MA-MAE}(\bm{l},\hat{\bm{l}}) = \frac{1}{C} \sum_{c=1}^{C} \frac{1}{n_c} \sum_{l_i \in T_c} |l_i-\hat{l_i}| \ ,
\end{equation}
where $C$ represents the number of classes, $n_c$ is the number of test samples in class $c$ and $T_c$ is the set of images in the test set whose true class label $l_i$ is $c$.

\subsection{Ranking scheme} %same as reubens paper
\subsubsection{Segmentation task} For the segmentation tasks, we adhered to the ranking scheme established in our first \crossmoda edition \cite{dorent2023crossmoda}, drawing inspiration from other renowned challenges such as BraTS \cite{bakas2018identifying} and ISLES \cite{maier2017isles}.  In this scheme, participating teams were ranked for each testing case, each evaluated region (VS and cochleas for the 2022 edition; and intrameatal, extrameatal and cochleas for the 2023 editions), and each metric (DSC and ASSD). For ties, the lowest rank among the tied values was assigned to each tied score. 

Rank scores were computed by first averaging the individual rankings for each case to obtain a cumulative rank. These cumulative ranks were then averaged across all patients for each participating team. Teams were ultimately ranked based on their final rank score. This ranking scheme was pre-defined, published before the challenge commenced, and made available on the Grand Challenge page and the \crossmoda website.

To assess the stability of this ranking scheme, we employed the bootstrapping method \cite{wiesenfarth2021methods}. We generated 1,000 bootstrap samples, each consisting of N=271 and N=341 test cases randomly drawn with replacement from the test sets used in the 2022 and 2023 editions, respectively, which also contained N=271 and N=341 cases. On average, each bootstrap sample retained approximately 63\% of distinct cases. The proposed ranking scheme was applied to each bootstrap sample.

The original ranking, calculated from the full test set, was then compared pairwise with the rankings derived from the bootstrap samples. We computed the correlation between these pairs of rankings using Kendall’s $\tau$, which ranges from -1 (indicating a completely reversed ranking order) to 1 (indicating an identical ranking order). This method allowed us to evaluate the robustness and stability of the ranking scheme employed in the challenge. 

% \paragraph{\crossmoda 2023 edition}
% For the 2023 edition, the intra-/extra-meatal sub-segmentation was not used for the ranking and instead combined intra-/extra-meal regions were used as the VS region.

\subsubsection{Koos classification tasks}
For the  Koos classification task, the MA-MAE score, which is computed over the testing dataset, is directly used to rank the participants.

\section{Participating methods}

\subsection{2021 edition}
% In our summary paper on the 2021 \crossmoda edition \citep{dorent2023crossmoda}, we detailed the number of participants in each phase and the domain adaptation techniques they employed.
In the 2021 \crossmoda edition, 55 teams from 16 different countries submitted predictions to the validation leaderboard. Among them, 16 teams from 9 different countries joined the evaluation phase. For a comprehensive overview of the methods proposed, we refer readers to our summary paper on the 2021 \crossmoda edition \citep{dorent2023crossmoda}.

\subsection{2022 edition}
In the 2022 \crossmoda edition, 233 teams from 35 countries registered for the challenge on the Grand Challenge platform. For the segmentation task, 27 teams submitted results to the validation leaderboard, with 12 teams from 8 countries joining the evaluation phase. For the classification task, 3 teams (colour coded in this manuscript as \textcolor{Super-Polymerization-c}{\Large\textbullet}, \textcolor{SJTU-EIEE-2-426Lab-c}{\Large\textbullet}, \textcolor{skjp-c}{\Large\textbullet}) participated in both the evaluation and testing phases. 

This section presents an overview of the methodologies employed for the DA task by the 12 participating teams. Each approach has been assigned a distinct colour code, consistently used across tables and figures for reference. A concise comparison of the methods is summarised in \tabref{tab:summary_2022}. 

%\clearpage
%\setlength\rotFPtop{20pt}
%\afterpage{%
%\begin{landscape}
%\begin{rotatepageccw}
%\begin{table*}[p]
\begin{sidewaystable*}[p]
%\begin{rotatepageccw}
\caption{Summary of the approaches used in the overall methodology, augmentation techniques and task parameters related to the segmentation task, and additional inference techniques for the submissions in the 2022 \crossmoda edition.
\label{tab:summary_2022}
}
\rotatepageforlabelcw{tab:summary_2022}
\begin{adjustbox}{scale=0.8,center}
\centering
\begin{tabular}{p{2.5cm}p{2cm}p{2.5cm}p{1cm}p{1.5cm}p{2.5cm}p{3.5cm}p{3cm}p{2cm}p{2.5cm}}
\hline
& \multicolumn{5}{c}{Methodology} & \multicolumn{2}{c}{Segmentation} & \multicolumn{2}{c}{Inference} \\\cmidrule(r){2-6}\cmidrule(rl){7-8}\cmidrule(l){9-10}
\textbf{Team} & \textbf{Image-to-image Translation Method} & \textbf{Segmentation Network} & \textbf{Self Training} & \textbf{Seg.-based Synthesis} & \textbf{Cropping} & \textbf{Data Augmentation} & \textbf{Parameters} & \textbf{Ensembling} & \textbf{Post-processing} \\
\hline
\textcolor{ne2e-c}{\Large\textbullet} ne2e & NiceGAN & 3D nnUNet & \checkmark   & \checkmark & Original image size dependent ROI & nnUNet augm. & nnUNet default &  5 $\times$ 3D nnUNet & VS:LCC\\ \hline
\textcolor{MAI-c}{\Large\textbullet} MAI & CycleGAN and QS-Attn & nnUNet & \checkmark & \xmark & center-cropping and resizing to fixed size & multi-view image translation + nnUNet augm. & nnUNet default & 5 $\times$ 3D nnUNet, 5 $\times$ 2D nnUNet & \xmark \\ \hline
\textcolor{LaTIM-c}{\Large\textbullet} LaTIM & CycleGAN & 3D nnUNet - full resolution & \checkmark & \xmark & Fixed ROI & structures: intensity scaling + SinGAN blending + nnUNet augm. & nnUNet default  & 5 * 3D nnUNet &  5 $\times$ 3D nnUNet, \\\hline
\textcolor{Super-Polymerization-c}{\Large\textbullet} Super Polymerization & MSF-Net & 3D nnUNet - full resolution & \checkmark   & \checkmark & Fixed ROI & nnUNet augm. & nnUNet default & 5 $\times$ 3D nnUNet & VS: LCC \\\hline
\textcolor{A*DA-c}{\Large\textbullet} A*DA & SE-CUT & 3D nnUNet - full resolution & \checkmark & \checkmark & Fixed ROI & structures: intensity augm. + nnUNet augm. & nnUNet default & 5-fold MS-MT models & VS: LCC cochlea: 2 LCC\\\hline
\textcolor{fgh-365-c}{\Large\textbullet} fgh\_365 & 2D and 3D CycleGAN & 3D nnUNet - full resolution & \checkmark & \xmark & Fixed ROI & structure: intensity augm. + oversampling small VS + nnUNet augm. & nnUNet default & ensembling 4-stage outputs & \xmark \\\hline
\textcolor{SJTU-EIEE-2-426Lab-c}{\Large\textbullet} SJTU\_EIEE\_2-426Lab & CycleGAN with ResNet & 3D nnUNet - full resolution & \checkmark & \xmark & \xmark & nnUNet augm. & nnUNet default & \xmark & \xmark \\\hline
\textcolor{MBZUAI-VS-c}{\Large\textbullet} MBZUAI\_VS & 2D CUT with StyleGAN backbone & 3D nnUNet - full resolution & \checkmark & \xmark & Fixed ROI & structures: intensity aug on pseudo T2, and other augmentation on real T2 + nnUNet augm. & nnUNet default + nnUNet variant with non-smooth Dice loss & ensemble of 2 models: default and non-smooth Dice loss & LCC \\
\hline
\textcolor{HUST-CBIB-c}{\Large\textbullet} HUST\_CBIB & CUT and CycleGAN & a hybrid convolutional network & \xmark & \xmark & Fixed ROI & generative augm. + VS: intensity augm. + flipping + rotation & Adam optimiser, learning-rate: 0.0002, 250 epochs & 5 fold ensembling & \xmark \\
\hline
\textcolor{skjp-c}{\Large\textbullet} SKJP & DCLGAN & 3D encoder-decoder networks  & \xmark & \xmark & Fixed ROI & intensity augm., patch sampling & adaptive t-vMF Dice loss, Adam optimiser, learning rate: 0.001, 300 epochs  & 5 fold ensembling & \xmark \\
\hline
\textcolor{gabybaldeon-c}{\Large\textbullet} gabybaldeon & CycleGAN, CUT, and StAC-DA & deeply supervised U-Net & \xmark & \xmark & generative augm. & - & soft Dice loss, Adam optimiser, learning rate: 0.0002, 200 epochs  5 fold ensembling & \xmark \\
\hline
\textcolor{Of-Men-and-Rabbits-c}{\Large\textbullet} Of\_Men\_and\_Rabbits & UVCGAN & nnUNet & \checkmark & \xmark & Fixed ROI & Rotations, blurring, mirroring, contrast and brightness adjustments & nnUNet default & \xmark & \xmark \\
\hline
\end{tabular}
\end{adjustbox}
\footnotesize{$^a$ Rotations, scaling, Gaussian noise, Gaussian blur, brightness, contrast, simulation of low resolution, gamma correction and mirroring \\ augm.: augmentation, LCC: Largest Connected Component, ROI: Region of Interest, }
%\end{rotatepageccw}
\end{sidewaystable*}
%\end{table*}
%\end{rotatepageccw}
%\end{landscape}
%}

\subsubsection*{\textcolor{ne2e-c}{\Large\textbullet} ne2e (Dong et al.) - 1st place} The authors presented an unsupervised cross-modality DA approach called PAST, which integrated pixel alignment and self-training to address domain shift between ceT1 and T2 modalities. Initially, labeled ceT1 scans were translated to the T2 domain using NiceGAN~\citep{chen2020reusing}, an enhanced CycleGAN variant that reused discriminators for encoding, improving training efficiency. These synthesised T2 scans were then employed to train a segmentation model via supervised learning. Compared to their previous PAST model used in \crossmoda 2021 \citep{dong2021unsupervised}, an extra segmentation network was introduced as part of the translation framework to better preserve intricate structures, as demonstrated by \cite{shin2022cosmos}. The self-training component further refined the segmentation model by iteratively generating and leveraging pseudo-labels from high-confidence pixels. The 3D segmentation task was performed using the nnUNet~\citep{isensee2021nnu} framework.

\subsubsection*{\textcolor{MAI-c}{\Large\textbullet} MAI (Kang et al.) - 2nd place} \citeauthor{MAIkang2022multi} proposed a multi-view image translation framework to generate pseudo T2 images from real ceT1 images, with diverse visual characteristics from different perspectives. This framework employed two distinct image translation models in parallel: CycleGAN~\citep{zhu2017unpaired} and QS-Attn~\citep{hu2022qs}, each offering distinct benefits. CycleGAN allowed for the faithful representation of the textures (or intensity) of ceT1 images, including the VS and cochlea, by leveraging its pixel-level cycle consistency loss. On the other hand, QS-Attn focused on preserving the structural integrity of ceT1 images by using patch-level contrastive loss, selecting domain-relevant features based on entropy, and performing contrastive learning to enhance structural preservation. Following the generation of pseudo T2 images by both CycleGAN and QS-Attn, nnUNet~\citep{isensee2021nnu} was trained as a segmentation model using the manual segmentations of the ceT1 scans. Finally, the trained model was used to generate pseudo-labels for real T2 images, which were iteratively used in a self-training process, updating the segmentation model with these pseudo-labeled real T2 images to achieve further performance enhancements.

\subsubsection*{\textcolor{LaTIM-c}{\Large\textbullet} LaTIM (Sall{\'e} et al.) - 3rd place} \citeauthor{sallé2023crossmodal} proposed a new data augmentation technique to expose segmentation networks to a greater diversity of tumour appearances and to circumvent the limitations of conventional cross-modal segmentation pipelines (i.e., synthesise, then segment) based on CycleGAN-like models. After CycleGAN training between modalities on \londonscgk only, ceT1 scans from both centres were forward passed to generate aligned pseudo T2 images. However, a domain gap between real and pseudo T2s persisted: 1) cochlea signal was attenuated (a common issue with CycleGANs as shown in \citep{cohen}, and 2) pseudo T2s lacked specific features from the Tilburg centre (for instance, large heterogeneously, highly textured VS). To circumvent these issues, tumour intensities in pseudo T2 scans were rescaled by a factor greater or lesser than unity (thus mimicking hyper or hypo signal). Altered tumours were then realistically blended back using a 2D one-shot generative SinGAN model that learned the distribution of a single pseudo T2 slice at multiple scales, allowing for tumour harmonisation on a slice-by-slice basis \citep{singan}. Iterative self-training of the augmented segmentation network (3D nnUNet backbone at full resolution) was then performed: a first model was trained using pseudo T2s and augmented pseudo T2s, then inferred on real T2 scans to obtain pseudo-labels on the target set. The same model was retrained iteratively using pseudo T2s, augmented pseudo T2s, and T2 sequences. After 3 iterations, the final segmentation network was used for testing. Hyperparameters of the SinGAN augmentation were empirically selected to cover the expected distribution based on pseudo-label statistics of the first pass of the segmentation model.

\subsubsection*{\textcolor{Super-Polymerization-c}{\Large\textbullet} Super Polymerization (Han et al.) - 4th place} \citeauthor{han2022unsupervised} proposed an unsupervised DA framework to address the challenges of cross-modality VS and cochlea segmentation, as well as Koos grade prediction. The framework is composed of three main components: DA, segmentation, and classification. The multi-sequence fusion network  (MSF-Net) for DA is designed to learn a shared latent representation that can be reconstructed into multiple modalities. Additionally, it employs proxy tasks, such as VS segmentation and brain parcellation, to maintain consistency in image structures during the DA process. After generating images in the target modality, the nnUNet model~\citep{isensee2021nnu} is used for segmentation with its default 3D full-resolution configuration. Instead of training the segmentation task with paired multiple modalities, the authors adopted a strategy of using the nnUNet model to predict different modalities independently, which increases the model's robustness and reduces the risk of overfitting. 

For Koos grade classification, they utilised a pre-trained encoder from the DA phase to incorporate prior knowledge from multiple modalities and leveraged the predicted segmentation map to highlight the region of interest. Additionally, they introduced both supervised~\citep{khosla2020supervised} and self-supervised contrastive learning approaches for pre-training to enhance classification performance. Supervised contrastive learning helps to increase the distance between subjects with different Koos grades, while self-supervised contrastive learning reduces the distance between modalities within the same subjects. The code is available as open source\footnote{\url{https://github.com/fiy2W/cmda2022.superpolymerization}}.

\subsubsection*{\textcolor{A*DA-c}{\Large\textbullet} A*DA (Zhao et al.) - 5th place}
The team proposed a Multi-Scale Mean Teacher (MS-MT) framework with segmentation-enhanced Contrastive Unpaired Translation (SE-CUT) for unsupervised cross-modality adaptation~\citep{zhao2023msmt}. First, the source T1 images were converted into synthetic T2 images using a modified 2D CUT with an additional segmentation decoder to preserve the structural information of the VS and cochlea. To generate pseudo labels for unlabeled T2 scans, a 3D full-resolution nnUNet \cite{isensee2021nnu} trained using pseudo T2, is utilised. To leverage both labelled pseudo T2 and unlabelled real T2 images, the team constructed a multi-scale self-ensembling network based on the mean teacher architecture, which provided deep supervision and consistency regularisation to further close the domain gap ~\citep{li2021hierarchical,zhao2022le}. Auxiliary layers were connected to each of the last five blocks in the segmentation network to obtain multi-scale predictions. To diversify the training samples of the segmentation network, intensity augmentation was performed by muting and intensifying the signal intensities of the VS and cochlea by 50\%, respectively. The segmentation backbone was a 3D full-resolution nnUNet, and post-processing was done by computing the largest connected component (LCC) to retain the first LCC for VS and the first two LCCs for cochlea.

\subsubsection*{\textcolor{fgh-365-c}{\Large\textbullet} fgh\_365 (Liu et al.) - 6th place} \citeauthor{liu2022enhancing} utilised a multi-step framework combining cross-site cross-modality image translation, rule-based offline augmentation, and iterative self-training for unsupervised segmentation of VS and cochlea. First, pseudo T2 images were generated from ceT1 images using CycleGANs, trained in five configurations to capture both intra-site and inter-site variations. Then the 3D nnUNet was trained with pseudo T2 generated from 2D and 3D CycleGANs (Stage 1). In Stage 2, those generated by 2D CycleGANs were replaced by real T2, enabling self-training to refine predictions. Stage 3 introduced offline augmentation, reducing VS intensities by 50\% and adjusting cochlea intensities to mimic real T2 characteristics. To improve the performance in small VS, pseudo T2 images with small tumours were also oversampled until the training dataset reached the nnUNet's limit of 1,000 samples. Finally, Stage 4 involved an additional round of self-training, with the final predictions generated through an ensemble model combining outputs from Stages 1, 2, and 3 using averaged softmax probabilities.  

\subsubsection*{\textcolor{SJTU-EIEE-2-426Lab-c}{\Large\textbullet} SJTU\_EIEE\_2-426Lab (Yang and Wang) - 7th place} \citeauthor{yang2025imagetrans} proposed an unsupervised cross-modality DA method for VS and cochlea segmentation that integrates image translation and self-training. Initially, real ceT1 scans were converted into the T2 modality to produce pseudo T2 scans using CycleGAN~\cite{zhu2017unpaired}. In this process, a residual neural network (ResNet) was employed as the generator instead of U-Net, while the default PatchGAN served as the discriminator. Subsequently, a segmentation model based on the nnUNet framework~\cite{isensee2021nnu} was trained on the pseudo T2 images along with their corresponding annotations. Finally, pseudo labels were generated for the real T2 scans, and the network was iteratively retrained on a combined dataset comprising pseudo T2 images (with real labels) and real T2 images (with generated pseudo labels). Notably, a single segmentation model was trained on all available data, given that the default 5-fold cross-validation of nnUNet combined with self-training can be very time-consuming. 

For the Koos classification task, the team built upon the current state-of-the-art methods for supervised Koos classification \citep{kujawa2022automated}. Specifically, they trained a nnUNet model to segment VS and seven surrounding brain structures involved in the Koos classification: the pons, brainstem, cerebellar vermal lobules I-V, VI-VII, and VIII-X, and the left and right cerebellum. Unlike  in \citeauthor{kujawa2022automated}, where pairs of T2 scans and annotations were available, the team used pseudo T2 scans along with their corresponding manual VS segmentations and automated GIF parcellations derived from ceT1 scans. The nnUNet framework was employed for 3D segmentation using the default full-resolution configuration. Additionally, hand-crafted features were extracted from both the brain structure segmentation masks and background masks, including volume, shortest distance to VS, and contact surface with VS. These features were then used to train a random forest classifier on the Koos classification training set, which was applied to predict Koos grades.

\subsubsection*{\textcolor{MBZUAI-VS-c}{\Large\textbullet} MBZUAI\_VS (Hardan et al.) - 8th place}
The team implemented a deep learning algorithm following a weak-supervision approach \cite{Hardan}. They began by resampling scans to an isotropic resolution of 1 $mm^3$, using third-order b-spline interpolation for images and nearest-neighbour for labels. Images were cropped or padded in the xy-plane to $256\times256$ and normalised on a 3D basis. The approach proceeded through two phases: domain translation and segmentation. In the domain translation phase, they used the 2D CUT method \cite{park2020contrastive} to transform ceT1 images into synthetic T2 images, replacing the original backbone with StyleGAN \cite{stylegan2} in the generator and discriminator networks. For segmentation, they executed three stages. First, a full-resolution 3D nnUNet \cite{isensee2021nnu} was trained using synthetic T2 images and their augmented versions with VS voxel values reduced by 50\%. Second, eight types of augmentations were applied to real T2 scans. The model from stage one generated segmentation masks for real T2 and their augmented versions. Third, images from the first stages were combined with stage 2 outputs to re-train the nnUNet. Lastly, final predictions are performed using an ensemble of two models: one model was trained with combined cross-entropy and Dice loss, while the other was trained with a non-smooth Dice loss. During inference, masks were post-processed by retaining the largest connected component of VS.

\subsubsection*{\textcolor{HUST-CBIB-c}{\Large\textbullet} HUST\_CBIB (Zhuang et al.) - 9th place}
The team proposed a two-stage unpaired cross-modality segmentation framework using data augmentation and hybrid convolutional networks \citep{zhuang2022unpaired}. Given the heterogeneous distributions and various voxel sizes between multi-institutional ceT1 and T2  scans, the framework first applied intensity normalisation, voxel size resampling, and centre cropping operations to obtain fixed-size sub-volumes for training. In the image synthesis stage, CUT and CycleGAN, allowed to align intensity distributions between unpaired scans \cite{liu2022bidirectional} and generate two groups of realistic and labelled T2 volumes with different details and appearances from real ceT1 volumes. For segmentation, the framework employed a hybrid convolutional network \cite{dong2022mnet} with nested architectures and multi-dimensional convolutions, learning to perform image segmentation from synthetic T2 volumes in a supervised manner. During training, they randomly adjusted VS tumour intensities between 0\% to 50\% in synthesised T2 scans for online augmentation to reduce overfitting risks and simulate tumour signal heterogeneity. At inference, the trained segmentation networks used multi-fold model ensembles and sliding window-based predictions to predict unlabeled T2 scans without additional post-processing.

\subsubsection*{\textcolor{skjp-c}{\Large\textbullet} SKJP (Kondo and Kasai) - 10th place} 
The team proposed a framework for cross-modality DA in three steps \citep{kondo2023unsupervised}. Initially, they translated ceT1 volumes into T2-like volumes using the DCLGAN method~\citep{han2021dual}, based on contrastive learning and a dual learning setting for efficient mapping between unpaired data. Translation was performed on 2D images, followed by segmentation on translated T2-like volumes using 3D encoder-decoder networks with a residual U-Net~\cite{futrega2022optimized} as the basic model. The loss function used was adaptive t-vMF Dice loss~\cite{kato2022adaptive}, incorporating deep supervision for loss computation. During training, 3D patches of size $96 \times 96 \times 96$ voxels were randomly sampled from input volumes, maintaining a 1:1:1 ratio of positive (VS and cochlea) to negative patches. Multiple models were trained independently using different combinations of training and validation datasets, with inference results obtained through ensemble outputs. 

The team also joined the Koos classification task. Koos classification was performed using linear SVM with hand-crafted features derived from VS segmentation results: the volume and size of the bounding box in x, y, and z directions.

\subsubsection*{\textcolor{gabybaldeon-c}{\Large\textbullet} gabybaldeon (Baldeon-Calisto et al.) - 11th place} The proposed framework consisted of two phases, an image-level adaptation phase followed by a feature-level adaptation phase. In phase one, source domain images were translated to the target domain using CycleGAN \cite{zhu2017unpaired}, CUT \cite{park2020contrastive}, and StAC-DA \cite{baldeon4075460stac} models. Given the unpaired nature of source and target domain images, each model performed a different mapping optimised by objective functions during training. Thus, after phase one, the translated source domain dataset was expanded threefold. In phase two, a deeply supervised U-Net was trained with a soft Dice loss, randomly selecting images and their corresponding ground-truth segmentations from the translated source dataset. Additionally, two discriminator networks were trained to distinguish target from source domain segmentations, while the U-Net aimed to deceive them. This adversarial learning approach facilitated feature-level adaptation by inducing the U-Net to learn domain-invariant features. To enhance robustness, five U-Net networks were trained using a five-fold division scheme and ensembled to produce final segmentation results.

\subsubsection*{\textcolor{Of-Men-and-Rabbits-c}{\Large\textbullet} Of\_Men\_and\_Rabbits (Noman et al.) - 12th place} This team implemented a two-phase framework with image translation and segmentation of VS and cochlea. In the first phase, UVCGAN \cite{torbunov2023uvcgan}, a CycleGAN variant using a Vision Transformer-based UNet (Vit-UNet) generator, was employed to translate ceT1 images into synthetic T2 images. In the second phase, nnUNet was used for segmentation, with offline data augmentation strategies, including rotation, blurring, Mirroring, contrast and brightness adjustments. The model was further fine-tuned through self-training, leveraging pseudo-labels generated from synthetic T2 images to improve performance on real T2 scans.

\begin{sidewaystable*}[p]
\caption{Summary of the approaches used in the overall methodology, augmentation techniques and task parameters related to the segmentation task, and additional inference techniques for the submissions in the 2023 \crossmoda edition.}
\label{tab:summary_2023}
%\rotatepageforlabelcw{tab:summary_2023}
\rotatepageforlabelcw{tab:summary_2023}
\begin{adjustbox}{scale=0.9,center}
\begin{tabular}{p{2.5cm}p{2cm}p{2.5cm}p{1cm}p{1.5cm}p{2.5cm}p{3.5cm}p{3cm}p{2cm}p{2.5cm}}
\hline
& \multicolumn{5}{c}{Methodology} & \multicolumn{2}{c}{Segmentation} & \multicolumn{2}{c}{Inference} \\ \cmidrule(r){2-6}\cmidrule(rl){7-8}\cmidrule(l){9-10}
\textbf{Team} & \textbf{Image-to-image Translation Method} & \textbf{Segmentation Network} & \textbf{Self Training} & \textbf{Seg.-based Synthesis} & \textbf{Cropping} & \textbf{Data Augmentation} & \textbf{Parameters} & \textbf{Ensembling} & \textbf{Post-processing} \\
\hline
\textcolor{vandy365-c}{\Large\textbullet} Vandy365 & 3D QS-Attn & 3D nnUNet - full resolution with U-Net and ResU-Net backbones & \checkmark & \checkmark & Cochlea based Fixed ROI & site-specific generative augm. + structures: intensity augm. + modified flipping and other nnUNet augm. + oversampling hard samples  & nnUNet default & 11 models with different configurations & CC \\
\hline
\textcolor{superpoly-c}{\Large\textbullet} SuperPolymerization & Seq2Seq generator and hyper discriminator & 3D nnUNet - full resolution  & \xmark & \xmark & Fixed ROI & site \& plane specific  generative augm.+ nnUNet augm. & nnUNet default & \xmark & - \\
\hline
\textcolor{ccc1018-c}{\Large\textbullet} ccc\_1018 & CUT & 2.5D ResUnet model & \checkmark & \xmark & Original image size dependent ROI & Random flipping, elastic transforms, contrast, brightness and gamma augmentation & Adam optimiser, learning rate 0.0002, 200 epochs & \xmark & aligning pseudo-labels by the intensity scaling factors \\
\hline
\textcolor{hustcbib-c}{\Large\textbullet} HUST\_CBIB & CUT, WCUT, 2.5D CycleGAN & 3D nnUNet & \checkmark & \checkmark & Fixed ROI & generative augm. + nnUNet augm. & nnUNet default & 3 fold ensembling & \xmark \\
\hline
\textcolor{auda-c}{\Large\textbullet} A-UDA & SE-CUT & 3D nnUNet - full resolution & \checkmark & \checkmark & Fixed ROI & generative augm. + structures: intensity augm. + nnUNet augm. & nnUNet default & 5-fold MS-MT models & VS: LCC cochlea: 2 LCC \\
\hline
\textcolor{mbzuai-c}{\Large\textbullet} MBZUAI-BioMedIA & CycleGAN & 3D nnUNet & \checkmark & \xmark & cropping
and resizing to
fixed size & nnUNet augm. & nnUNet default & 3 fold ensembling & filling the gaps \& removing isolated predictions.\\
\hline
\end{tabular}
\end{adjustbox}
\footnotesize{$^a$ Rotations, scaling, Gaussian noise, Gaussian blur, brightness, contrast, simulation of low resolution, gamma correction and mirroring \\ augm.: augmentation, LCC: Largest Connected Component, ROI: Region of Interest, }
\end{sidewaystable*}

\subsection{2023 edition}
Out of 26 registered teams, 13 teams from 7 different countries participated in the validation leaderboard, while 6 teams from 5 different countries participated in the testing leaderboard.

An overview of the key aspects of each approach followed by the 6 teams is presented in \tabref{tab:summary_2023}, providing a concise comparison of their methodologies.

%\clearpage
%\vspace*{200pt}
%\afterpage{
%\begin{rotatepageccw}

%\end{landscape}
%\end{rotatepageccw}
%}

%\clearpage
%

\begin{figure*}[!ht]
    \centering
    \begin{subfigure}{0.33\textwidth}
            \centering
            \includegraphics[width=\linewidth]{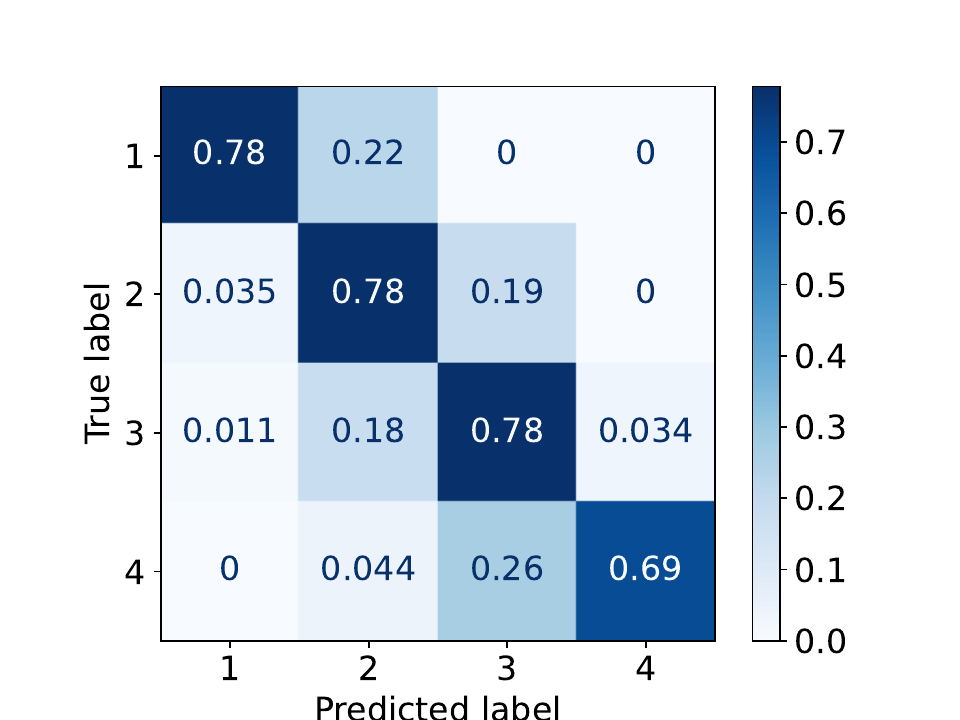}
            \caption{1st: \textcolor{SJTU-EIEE-2-426Lab-c}{\Large\textbullet} SJTU\_EIEE\_2-426Lab}
            \label{fig:confusion_matrix_koos_SJTU}
        \end{subfigure}
    \begin{subfigure}{0.33\textwidth}
        \centering
        \includegraphics[width=\linewidth]{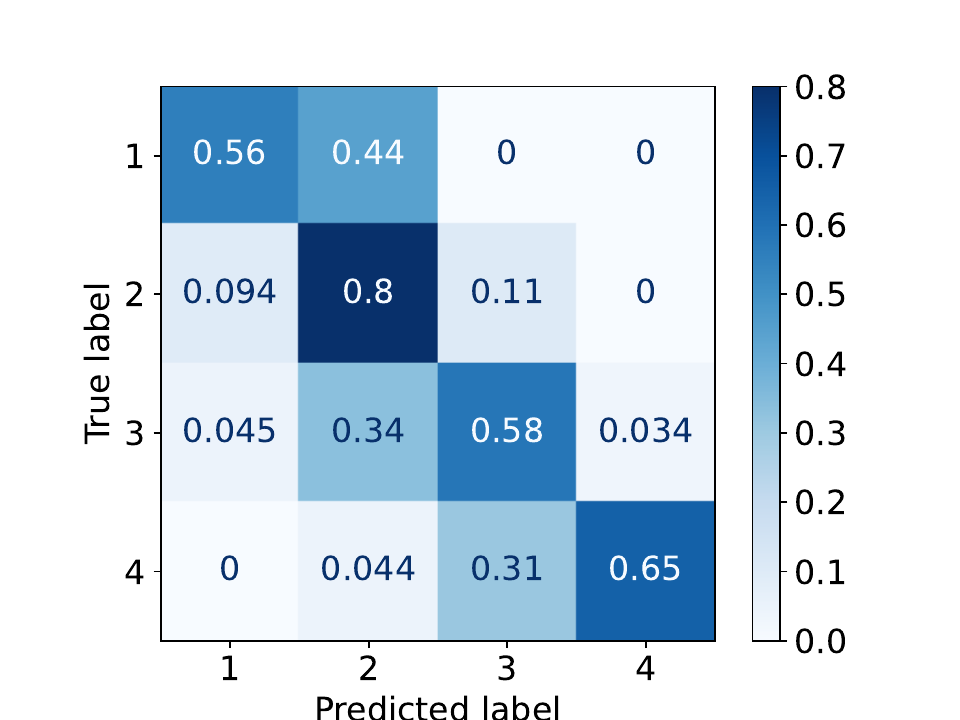}
        \caption{2nd: \textcolor{Super-Polymerization-c}{\Large\textbullet} Super Polymerization}
        \label{fig:confusion_matrix_spolym}
    \end{subfigure}
    \hfill
    \begin{subfigure}{0.33\textwidth}
        \centering
        \includegraphics[width=\linewidth]{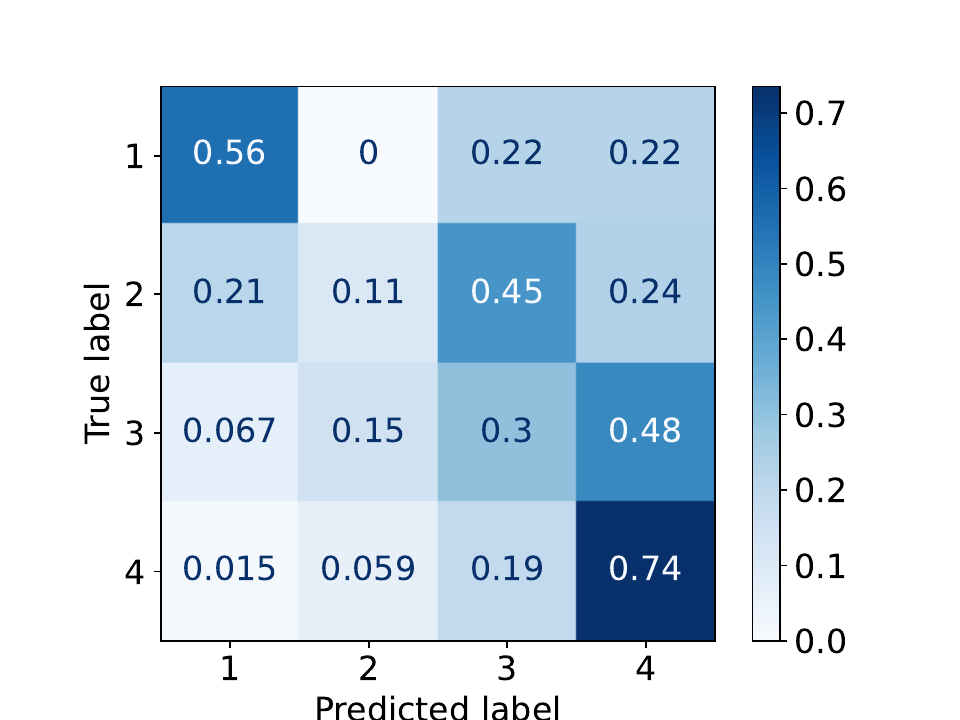}
        \caption{3rd: \textcolor{skjp-c}{\Large\textbullet} skjp}
        \label{fig:confusion_matrix_koos_skjp}
    \end{subfigure}
    \caption{Normalised confusion matrices illustrating the performance of the teams in the Koos classification task during the \crossmoda 2022 challenge.}

    \label{fig:confusion_matrices_koos}
\end{figure*}

\subsubsection*{\textcolor{vandy365-c}{\Large\textbullet} Vandy365 (Liu Han et al.) - 1st place}  
\citeauthor{liu2023learning} thoroughly analysed the approaches followed in previous editions and derived solutions from them. They addressed the challenge of multi-institutional unsupervised DA in medical imaging by following the commonly used three-step strategy focused on image-level domain gap reduction. For preprocessing, they have used a 2D QS-Attn \cite{hu2022qs} based synthesis followed by a nnUNet based segmentation approach to localise the cochlea for cropping purposes. In step 1, unpaired image translation is used to convert ceT1 images into synthetic T2 images using extended 3D QS-Attn with label-assisted intensity transformation to ensure anatomical consistency. This synthesis network consists of an extra segmentation decoder that could make the generator focus more on the anatomical structures. Moreover, this network can generate a broader spectrum of synthetic data by synthesizing site-specific T2 styles through dynamic instance normalization \cite{liu2023learning,liu2022moddrop++}.  In step 2, these synthetic T2 images and the associated ceT1 labels are used to train the 3D full-resolution nnUNet segmentation model. In step 3, the method further reduces the domain gap by using unlabeled real T2 images to train the segmentation model via self-training, enriching the training dataset with diverse synthetic styles and oversampling hard samples to improve segmentation performance. Lastly, they used a model ensemble by averaging the predictions from 11 models to further boost the performance. These models include 3 standard nnUNet models trained with different seeds and 8 customised nnUNet models with the following configurations: 2 different backbone architectures (U-Net or ResU-Net) × 2 different augmentation strategies (strong or weak local intensity augmentation for VS and cochleae) × 2 different sets of unseen site codes for style interpolation.  

\subsubsection*{\textcolor{ccc1018-c}{\Large\textbullet} ccc\_1018 (Can Can et al.) 2nd place} This proposed method is built on established deep learning frameworks for unpaired image translation and semantic segmentation. For image synthesis in step 1, the CUT model \cite{park2020contrastive} is employed for unpaired image-to-image translation from ceT1 to T2 scans, leveraging patch-wise contrastive learning and adversarial learning, with a generator and discriminator setup, optimising the process with specific hyper-parameters and instance normalisation. For segmentation in step 2, the 2.5D ResUnet model \cite{diakogiannis2020resunet} is used with a high-performance configuration that includes three down-sample and three up-sample layers, without z-direction down-sampling. The segmentation process involves the step 3 of self-training, where the initial model is trained on translated T2 images, pseudo-labels are generated and post-processed, and the model is iteratively retrained. %Data augmentation includes hard example mining and adjustments for image heterogeneity by randomly replacing or expanding image sections.

\subsubsection*{\textcolor{superpoly-c}{\Large\textbullet} SuperPolymerization (Luyi Han et al.) - 3rd place} This team's approach to cross-modality DA leverages a fine-grained unsupervised framework to enhance segmentation accuracy for VS and cochlea. The image synthesis stage of this framework utilises Seq2Seq generator, which synthesises T2 images from real ceT1 images by controlling features through conditional codes \cite{han2024synthesis}. This process generates diverse synthetic images by varying factors such as the image center and plane view, effectively augmenting the dataset with nine unique versions of each real ceT1 image. Additionally, they proposed a hyperdiscriminator in this DA framework, which can differentiate between true and false images, under different conditional codes. The augmented dataset, enriched with a variety of synthetic T2 images, is subsequently used to train the nnUNet \cite{isensee2021nnu} model for Segmentation. 

\subsubsection*{\textcolor{hustcbib-c}{\Large\textbullet} HUST\_CBIB (Zhuang et al.) - 4th place}  This team focused on generating realistic target-like volumes from labeled source scans using multi-style image translation strategies. They trained 2.5D CycleGAN with cycle-consistency loss ($G_{CycleGAN}$) \cite{zhu2017unpaired}, CUT with contrastive loss ($G_{CUT}$) \cite{park2020contrastive}, 2D weighted contrastive unpaired image translation network (WCUT) ($G_{WCUT}$) to produce high-quality target-like volumes each with an auxiliary segmentation loss to enhance the translation performance in segmentation regions. In the second stage, the team employed an iterative self-training method for 3D segmentation using the nnUNet framework \cite{isensee2021nnu} from labeled fake T2 volumes obtained with $G_{CycleGAN}$, $G_{CUT}$ and $G_{WCUT}$. The process involved initially training the segmentation model on labeled fake T2 volumes, generating pseudo-labels for unlabeled real T2 volumes, and retraining the model with a combination of labeled fake T2 and pseudo-labeled real T2 volumes. This iterative retraining aimed to reduce the distribution gap between synthetic and real T2 images, thereby improving the segmentation model's robustness and performance on unseen real T2 scans.

\subsubsection*{\textcolor{auda-c}{\Large\textbullet} A-UDA (Zhao et al.) - 5th place}
This approach begins by translating ceT1 scans to the T2 modality using a SE-CUT network. This network builds upon the 2D CUT \cite{park2020contrastive} method and enhances its performance with an additional segmentation decoder, which improves the preservation of structural details during the translation process. Three CycleGAN networks are then employed for pixel-level intensity fine-tuning. They also perform intensity augmentation on the annotated regions of the generated images to diversify the training data. These synthetic and augmented scans are used for training a 3D nnUNet, which can generate pseudo labels for all unlabeled real T2 images \cite{isensee2021nnu}. Finally, a MS-MT network is constructed for self-training, leveraging a teacher-student model updated via exponential moving average to ensure consistency and deep supervision through multi-scale predictions. 

\subsubsection*{\textcolor{mbzuai-c}{\Large\textbullet} MBZUAI-BioMedIA (Santosh Sanjeev et al.) - 6th place}
This team explores several techniques for image translation from ceT1 to T2 scans using three distinct methods: pixel-level consistency with CycleGAN \cite{zhu2017unpaired}, patch-level contrastive constraints through QS-Attn, and density constraints via DECENT \cite{xie2022unsupervised}. However, they employed the traditional 2D CycleGAN for image translation. These translated T2 scans are then utilised to train a 3D segmentation model, nnUNet \cite{isensee2021nnu}, for accurate tumour identification and segmentation. The authors further refine the segmentation process through self-training, where the model iteratively learns from pseudo-labeled T2 scans and employ postprocessing techniques to address issues such as isolated predictions and systematic errors in tumour localisation, thereby improving the overall segmentation quality.

% \FloatBarrier

\section{Results}
In this section, we present the key results from the 2022 and 2023 \crossmoda challenges and provide a retrospective analysis of the challenge's evolution over the years. We begin with the Koos classification task from 2022, followed by an in-depth analysis of the segmentation task, which attracted the highest level of participation.

\subsection{Koos classification task from 2022 edition}

In 2022, 3 teams participated in the Koos classification task. Participants were required to submit their Docker containers by August 15, 2022, and the winners were announced during the \crossmoda event at the MICCAI 2022 conference.

The final scores and rankings are presented in \tabref{tab:ma_mae_Koos}. The top-performing team (\textcolor{SJTU-EIEE-2-426Lab-c}{\Large\textbullet} SJTU\_EIEE\_2-426Lab) achieved a MA-MAE score of 0.26, which is significantly higher than the intra-rater variability, estimated at 0.11 \citep{kujawa2022automated}.   
The confusion matrices for the top three teams are shown in \figref{fig:confusion_matrices_koos}. Unlike the second- and third-ranked teams, the majority of misclassifications by the top-performing team were off by only a single grade level, with just 7 out of 251 cases misclassified by two grades. Notably, most errors occurred between Koos grades 3 and 4. Due to the limited number of participants, this task was discontinued in the 2023 edition.

\begin{table}[h!]
\centering
\caption{Classification ranking and corresponding MA-MAE scores for \crossmoda 2022. Arrows indicate favourable direction of the MA-MAE metric.}
\label{tab:ma_mae_Koos}
\begin{tabular}{lcc}
\toprule
&\multicolumn{1}{c}{Challenge rank} & \multicolumn{1}{c}{Koos Classification} \\\cmidrule{2-2}\cmidrule{3-3}
& Global rank & MA-MAE $\downarrow$\\ 
\midrule
\textcolor{SJTU-EIEE-2-426Lab-c}{\Large\textbullet} SJTU\_EIEE... & 1 & 0.26 \\
\rowcolor{LightGray}
\textcolor{Super-Polymerization-c}{\Large\textbullet} Super\_Poly... & 2 & 0.37\\
\textcolor{skjp-c}{\Large\textbullet} SKJP & 3 & 0.84 \\ \bottomrule
\end{tabular}
\end{table}

\subsection{Overall segmentation performance 2021 edition}

The overall segmentation performance of the participating teams in the 2021 \crossmoda edition is extensively discussed in our previous \crossmoda summary study \citep{dorent2023crossmoda}. 

\begin{figure*}[tbp!]
\begin{center}
\includegraphics[width=0.7\textwidth]{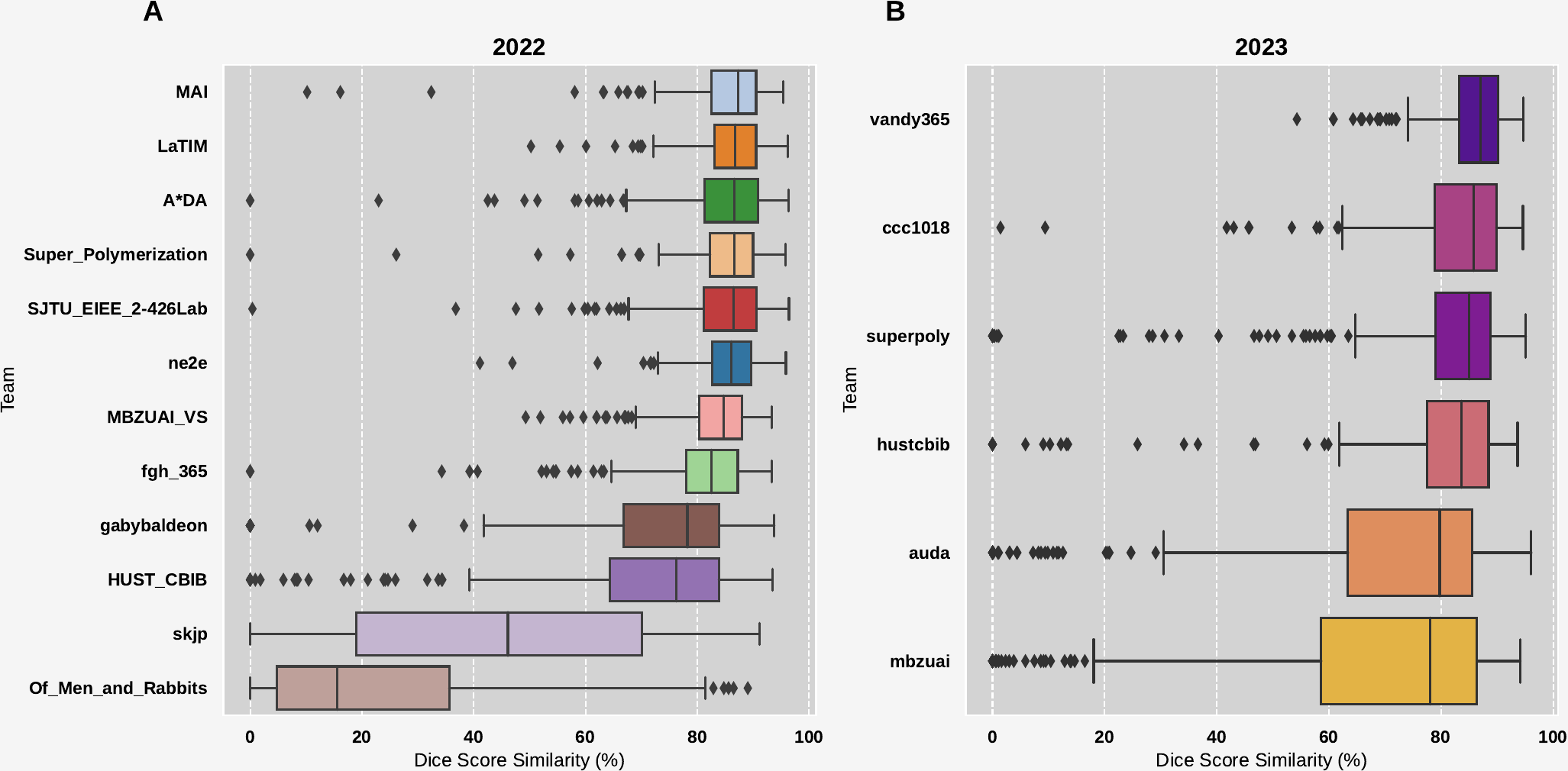}
\end{center}
\caption{Box plot of the methods' segmentation performance for the VS structure in terms of DSC for \textbf{(A)} 2022 edition \textbf{(B)} 2023 edition.}\label{fig:vs_dice}
\end{figure*}
\begin{figure*}[tbp!]
\begin{center}
\includegraphics[width=0.7\textwidth]{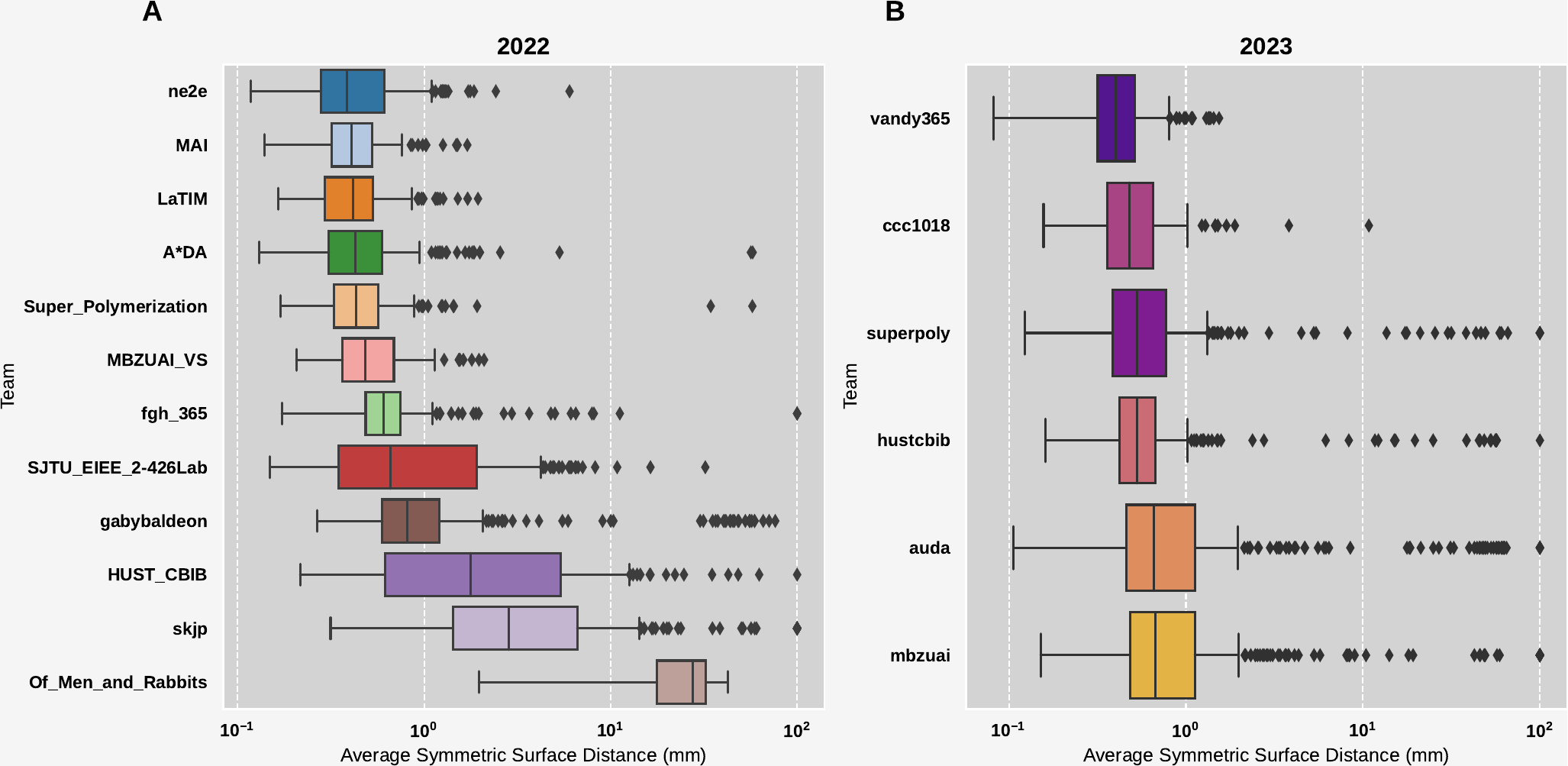}
\end{center}
\caption{Box plot of the methods' segmentation performance for the VS structure in terms of ASSD for \textbf{(A)} 2022 edition \textbf{(B)} 2023 edition.}\label{fig:vs_assd}
\end{figure*}

\begin{figure*}[tbp!]
\begin{center}
\includegraphics[width=0.7\textwidth]{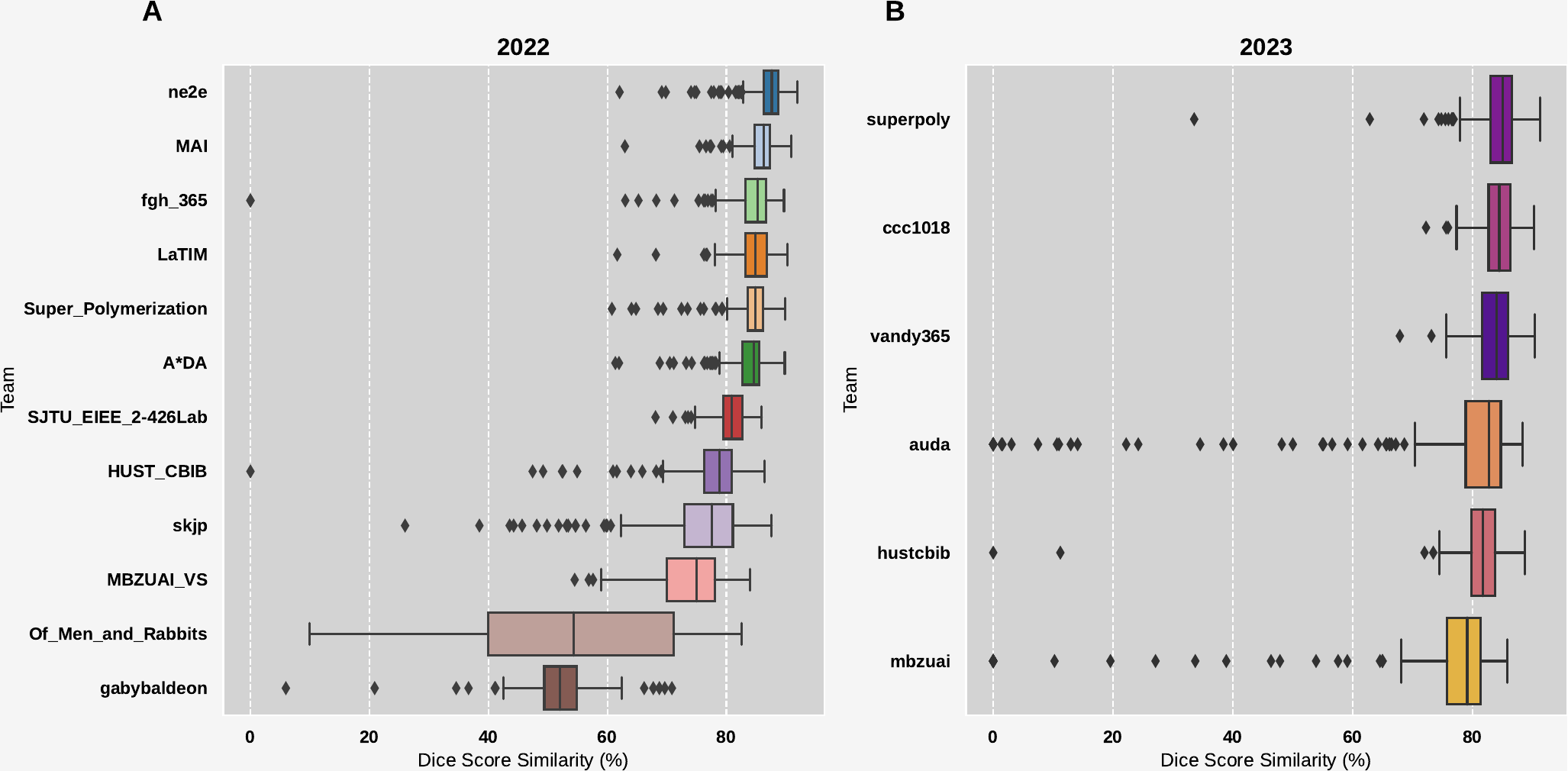}
\end{center}
\caption{Box plot of the methods' segmentation performance for the cochlea in terms of DSC for \textbf{(A)} 2022 edition \textbf{(B)} 2023 edition.}\label{fig:cochlea_dice}
\end{figure*}

\begin{figure*}[tbp!]
\begin{center}
\includegraphics[width=0.7\textwidth]{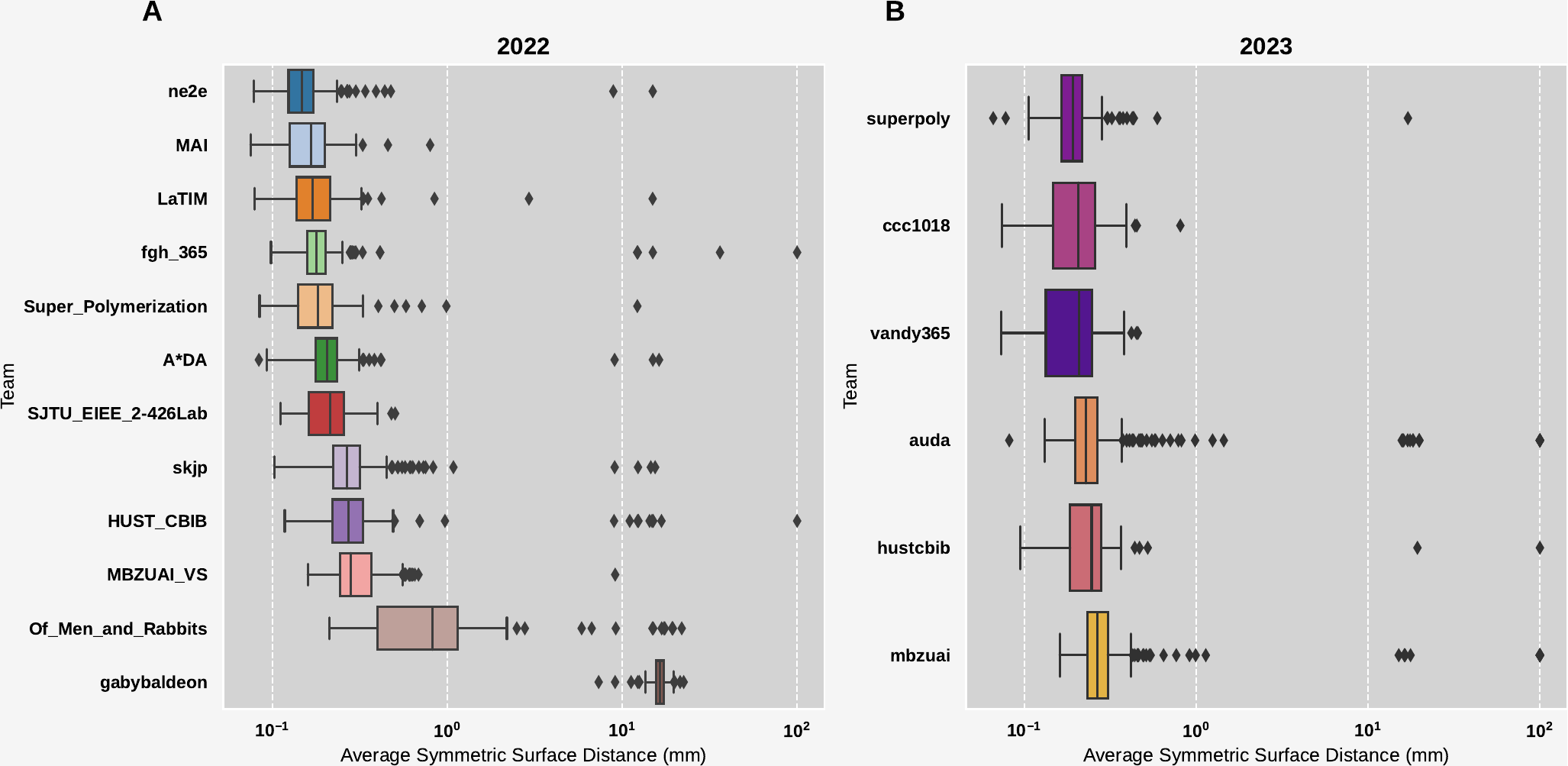}
\end{center}
\caption{Box plot of the methods' segmentation performance for the cochlea in terms of ASSD for \textbf{(A)} 2022 edition \textbf{(B)} 2023 edition.}\label{fig:cochlea_assd}
\end{figure*}

\begin{sidewaystable*}[p]
\caption{Segmentation metrics and corresponding scores for each submission, including median and interquartile range values for \crossmoda 2022 and 2023 editions. Arrows indicate favourable direction of each metric.}
\label{tab:dsc_assd_distribution}
\rotatepageforlabelccw{tab:dsc_assd_distribution}
\begin{adjustbox}{scale=0.9,center}
\begin{tabular}{lp{1cm}p{1cm}cccccc}
\toprule
&\multicolumn{2}{c}{Challenge rank} & \multicolumn{4}{c}{Vestibular Schwannoma} &\multicolumn{2}{c}{Cochlea} \\\cmidrule(r){2-3}\cmidrule(rl){4-5}\cmidrule(l){6-7}
& Global rank & Rank score & \multicolumn{2}{c}{DSC (\%) $\uparrow$} & \multicolumn{2}{c}{ASSD (mm) $\downarrow$} & DSC (\%)  $\uparrow$& ASSD (mm) $\downarrow$\\ 
\midrule
\textcolor{ne2e-c} {\Large\textbullet}  ne2e & 1 & 3.0 & \multicolumn{2}{c}{86.1 [82.7 - 89.7]} & \multicolumn{2}{c}{0.38 [0.28 - 0.61]} & 87.6 [86.3 - 88.7] & 0.15 [0.12 - 0.17] \\ 

\rowcolor{LightGray}
\textcolor{MAI-c} {\Large\textbullet}  MAI & 2 & 3.4 & \multicolumn{2}{c}{87.3 [82.5 - 90.5]} & \multicolumn{2}{c}{0.41 [0.32 - 0.53]} & 86.2 [84.8 - 87.3] & 0.17 [0.12 - 0.2] \\ 

\textcolor{LaTIM-c} {\Large\textbullet}  LaTIM & 3 & 3.8 & \multicolumn{2}{c}{86.8 [83.1 - 90.5]} & \multicolumn{2}{c}{0.42 [0.29 - 0.53]} & 84.9 [83.2 - 86.8] & 0.17 [0.14 - 0.21] \\ 

\rowcolor{LightGray}
\textcolor{Super-Polymerization-c} {\Large\textbullet}  Super\textunderscore Polymerization & 4 & 4.3 & \multicolumn{2}{c}{86.6 [82.3 - 90.0]} & \multicolumn{2}{c}{0.43 [0.33 - 0.57]} & 84.9 [83.6 - 86.2] & 0.18 [0.14 - 0.22] \\ 

\textcolor{A*DA-c} {\Large\textbullet}  A*DA & 5 & 4.9 & \multicolumn{2}{c}{86.7 [81.3 - 90.9]} & \multicolumn{2}{c}{0.43 [0.31 - 0.59]} & 84.6 [82.6 - 85.5] & 0.2 [0.18 - 0.23] \\ 

\rowcolor{LightGray}
\textcolor{fgh-365-c} {\Large\textbullet}  fgh\textunderscore 365 & 6 & 5.6 & \multicolumn{2}{c}{82.5 [78.0 - 87.2]} & \multicolumn{2}{c}{0.61 [0.49 - 0.75]} & 85.3 [83.3 - 86.6] & 0.18 [0.16 - 0.2] \\ 

\textcolor{SJTU-EIEE-2-426Lab-c} {\Large\textbullet}  SJTU\textunderscore EIEE\textunderscore 2-426Lab & 7 & 5.9 & \multicolumn{2}{c}{86.4 [81.1 - 90.6]} & \multicolumn{2}{c}{0.66 [0.35 - 1.91]} & 80.9 [79.5 - 82.7] & 0.21 [0.16 - 0.26] \\ 

\rowcolor{LightGray}
\textcolor{MBZUAI-VS-c} {\Large\textbullet}  MBZUAI\textunderscore VS & 8 & 7.5 & \multicolumn{2}{c}{84.7 [80.3 - 88.0]} & \multicolumn{2}{c}{0.49 [0.37 - 0.69]} & 75.0 [70.0 - 78.0] & 0.28 [0.24 - 0.37] \\ 

\textcolor{HUST-CBIB-c} {\Large\textbullet}  HUST\textunderscore CBIB & 9 & 8.6 & \multicolumn{2}{c}{76.3 [64.3 - 83.9]} & \multicolumn{2}{c}{1.78 [0.62 - 5.42]} & 78.9 [76.2 - 80.9] & 0.27 [0.22 - 0.33] \\ 

\rowcolor{LightGray}
\textcolor{skjp-c} {\Large\textbullet}  skjp & 10 & 9.5 & \multicolumn{2}{c}{46.1 [19.0 - 70.1]} & \multicolumn{2}{c}{2.83 [1.43 - 6.62]} & 77.6 [73.0 - 81.1] & 0.26 [0.22 - 0.31] \\ 

\textcolor{gabybaldeon-c} {\Large\textbullet}  gabybaldeon & 11 & 10.3 & \multicolumn{2}{c}{78.2 [66.9 - 83.9]} & \multicolumn{2}{c}{0.81 [0.6 - 1.2]} & 52.0 [49.4 - 54.8] & 16.44 [15.63 - 17.27] \\ 

\rowcolor{LightGray}
\textcolor{Of-Men-and-Rabbits-c} {\Large\textbullet}  Of\textunderscore Men\textunderscore and\textunderscore Rabbits & 12 & 11.3 & \multicolumn{2}{c}{15.6 [4.8 - 35.6]} & \multicolumn{2}{c}{27.61 [17.72 - 32.39]} & 54.4 [40.0 - 71.1] & 0.82 [0.4 - 1.14] \\ 

\midrule
&\multicolumn{2}{c}{Challenge rank} & \multicolumn{2}{c}{Intra-meatal VS} &\multicolumn{2}{c}{Extra-meatal VS} &\multicolumn{2}{c}{Cochlea} \\\cmidrule(r){2-3}\cmidrule(rl){4-5}\cmidrule(l){6-7}\cmidrule(l){8-9}
& Global rank & Rank score & DSC$\uparrow$ & ASSD (mm) $\downarrow$& DSC$\uparrow$ & ASSD (mm) $\downarrow$& DSC (\%)  $\uparrow$& ASSD (mm) $\downarrow$\\ \midrule
\textcolor{vandy365-c} {\Large\textbullet} vandy365 & 1 & 2.2 & 74.9 [68.8 - 80.0] & 0.43 [0.32 - 0.57] & 87.3 [82.7 - 90.4] & 0.4 [0.32 - 0.52] & 84.1 [81.7 - 86.0] & 0.21 [0.13 - 0.25] \\

\rowcolor{LightGray}
\textcolor{ccc1018-c} {\Large\textbullet} ccc1018 & 2 & 2.9 & 71.7 [64.5 - 76.9] & 0.54 [0.4 - 0.7] & 84.9 [75.9 - 89.4] & 0.48 [0.37 - 0.65] & 84.5 [82.7 - 86.3] & 0.21 [0.15 - 0.26] \\

\textcolor{superpoly-c} {\Large\textbullet} superpoly & 3 & 3.0 & 67.2 [53.0 - 76.8] & 0.66 [0.41 - 1.01] & 85.6 [77.2 - 89.4] & 0.49 [0.37 - 0.67] & 85.1 [83.1 - 86.6] & 0.19 [0.16 - 0.22] \\

\rowcolor{LightGray}
\textcolor{hustcbib-c} {\Large\textbullet} hustcbib & 4 & 3.5 & 69.7 [59.8 - 76.3] & 0.57 [0.4 - 0.78] & 83.5 [75.8 - 88.7] & 0.51 [0.43 - 0.64] & 81.8 [79.9 - 83.8] & 0.25 [0.18 - 0.28] \\

\textcolor{auda-c} {\Large\textbullet} auda & 5 & 4.2 & 67.1 [46.4 - 76.8] & 0.64 [0.42 - 1.21] & 78.2 [60.0 - 85.3] & 0.66 [0.48 - 1.01] & 82.7 [78.9 - 84.7] & 0.23 [0.2 - 0.27] \\

\rowcolor{LightGray}
\textcolor{mbzuai-c} {\Large\textbullet} mbzuai & 6 & 4.9 & 63.9 [46.7 - 72.8] & 0.72 [0.48 - 1.15] & 76.3 [51.2 - 85.8] & 0.67 [0.51 - 1.15] & 79.1 [75.8 - 81.4] & 0.27 [0.23 - 0.31] \\
\bottomrule
\end{tabular}
\end{adjustbox}
\end{sidewaystable*}

\subsection{Overall segmentation performance 2022 edition}
Participants were required to submit their Docker containers by 15th August 2022. Winners were announced during the \crossmoda event at the MICCAI 2022 conference. This section presents the results obtained by participant teams on the test set and analyses the stability and robustness of the proposed ranking scheme.

The final ranking for the 12 teams in the \crossmoda 2022 challenge is presented in \tabref{tab:dsc_assd_distribution}, organised by rank. \figref{fig:vs_dice} (A) and \figref{fig:vs_assd} (A) visualise box plots, colour-coded by team, for the VS structure using the DSC and ASSD metrics, respectively. Similarly, \figref{fig:cochlea_dice} (A) and \figref{fig:cochlea_assd} (A) show box plots, also colour-coded by team, for the cochlea structure, using the DSC and ASSD metrics.

The top-ranked team, \textit{ne2e}, achieved a rank score of 3.0, followed by \textit{MAI} with a rank score of 3.4. Both teams achieved a median DSC above 86\% for both cochlea and VS segmentation. Teams \textit{LaTIM}, \textit{Super Polymerization}, and \textit{A*DA} also performed well, with median DSCs exceeding 86\% for the VS structure and 84\% for the cochlea structure. In contrast, lower-ranking teams exhibited lower DSC values and higher ASSD values, highlighting the challenges of cross-modality DA with multi-institutional data.

All participating teams employed image-to-image translation approaches, predominantly using CycleGAN and its extensions, to bridge the gap between T1 and T2 scans.

\subsection{Overall segmentation performance 2023 edition}

Participants were required to submit their Docker containers by 10th July 2023. Winners were announced during the \crossmoda event at the MICCAI 2023 conference. This section presents the results obtained by participant teams on the test set and analyses the stability and robustness of the proposed ranking scheme.

The final ranking for the six teams in the \crossmoda 2023 challenge is presented in \tabref{tab:dsc_assd_distribution}, organised by rank. \figref{fig:vs_dice} (B) and \figref{fig:vs_assd} (B) visualise box plots, colour-coded by team, for the VS structure, based on the DSC and ASSD metrics, respectively. Similarly, \figref{fig:cochlea_dice} (B) and \figref{fig:cochlea_assd} (B) visualise box plots, also colour-coded by team, for the cochlea structure, using the DSC and ASSD metrics.

The top-ranked team, \textit{vandy365}, achieved a rank score of 2.3, with the highest median DSC (over 87\%) for the VS segmentation task and the second-highest median DSC for the cochlea segmentation. Teams \textit{superpoly} and \textit{ccc1018} achieved the same rank score of 2.8, with \textit{superpoly} achieving the second highest median DSC for the cochlea segmentation task. \textit{ccc1018} scored second highest for VS structure, with median DSCs above 85\%.

\begin{figure*}[tbp!]
\begin{center}
\includegraphics[width=0.74\textwidth]{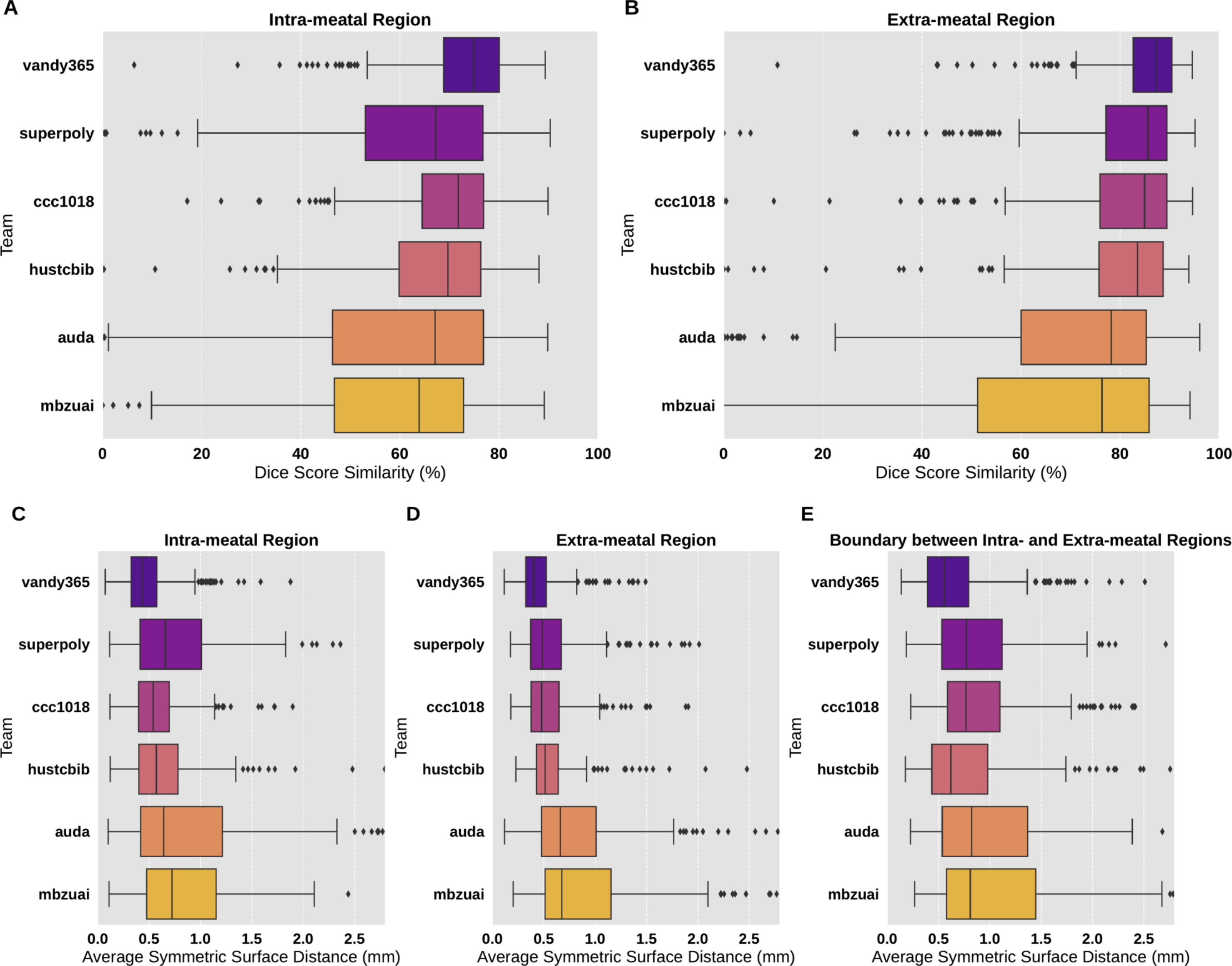}
\end{center}
\caption{Box plot of the methods' intra-/extra-meatal segmentation performance in 2023 edition in terms of DSC for \textbf{(A)} intra-meatal region\textbf{(B)} extra-meatal region and in terms of ASSD for \textbf{(C)} intra-meatal region \textbf{(D)} extra-meatal region \textbf{(E)} boundary separating intra-/extra- regions. }\label{fig:intra_extra_dice_assd}
\end{figure*}

\paragraph{Intra- / extra-meatal segmentation}
%
% Although intra- / extra-meatal segmentation performance was not considered for ranking purposes, we use this subtask to provide additional insight.
%
The top-ranked team, \textit{vandy365}, achieved the highest DSC for both intra- and extra-meatal regions, with a DSC score exceeding 87\% for the extra-meatal region and over 74\% for the intra-meatal region. \textit{ccc1018} achieved the second-highest intra-meatal DSC, exceeding 71\%, and the third-highest DSC, above 84\%, for the extra-meatal region. \textit{superpoly} recorded the third-highest median DSC, scoring over 67\% for the intra-meatal segmentation task, with median DSCs above 85\% for the extra-meatal component. Overall, the VS tumour DSC is slightly higher for \textit{ccc1018}, highlighting the contribution of the intra-meatal component to the overall DSC of the VS structure. \figref{fig:intra_extra_dice_assd} (A) and \figref{fig:intra_extra_dice_assd} (B) visualise DSC distribution box plots, colour-coded by team, for the intra- and extra-meatal regions respectively.
The top-ranked team, \textit{vandy365}, also achieved the lowest ASSD for the intra- and extra-meatal regions, as well as for the boundary delineation. 
The DSC and ASSD scores for the intra-/extra-meatal segmentation task are presented in \tabref{tab:dsc_assd_distribution}, organised by rank. \figref{fig:intra_extra_dice_assd}~(C) and \figref{fig:intra_extra_dice_assd}~(D) shows ASSD metric distribution box plots, colour-coded by team, for the intra-/extra-meatal regions and the split boundary respectively. \tabref{tab:intra_extra_performance} presents additional VS segmentation performance results, including split boundary ASSD metric that were provided to participants for reference but were not used in the final ranking.

\begin{table*}[htb!]
\centering
\caption{Additional metrics and corresponding scores for the VS segmentation task in \crossmoda 2023 edition, including median and interquartile range values. Arrows indicate favourable direction of each metric.}
\label{tab:intra_extra_performance}
\begin{tabular}{lccc}\\ \toprule
\multirow{2}{*}{\textbf{Team}} & \multicolumn{2}{c}{\textbf{VS}} & \textbf{Split Boundary} \\  \cmidrule(r){2-3}%\cmidrule(l){4-5}
& \textbf{DSC (\%) $\uparrow$} & \textbf{ASSD (mm) $\downarrow$}& \textbf{ASSD (mm) $\downarrow$} \\\midrule     
\textcolor{vandy365-c} {\Large\textbullet} vandy365 & 87.1 [83.3 - 90.2] & 0.4 [0.32 - 0.51] & 0.55 [0.39 - 0.79] \\

\rowcolor{LightGray}
\textcolor{ccc1018-c} {\Large\textbullet} ccc1018 & 85.8 [78.8 - 89.9] & 0.48 [0.36 - 0.65] & 0.77 [0.59 - 1.1] \\

\textcolor{superpoly-c} {\Large\textbullet} superpoly & 85.0 [79.0 - 88.8] & 0.53 [0.39 - 0.77] & 0.77 [0.53 - 1.12] \\

\rowcolor{LightGray}
\textcolor{hustcbib-c} {\Large\textbullet} hustcbib & 83.6 [77.5 - 88.5] & 0.53 [0.42 - 0.67] & 0.62 [0.43 - 0.98] \\

\textcolor{auda-c} {\Large\textbullet} auda & 79.8 [63.4 - 85.6] & 0.66 [0.46 - 1.13] & 0.82 [0.54 - 1.37] \\

\rowcolor{LightGray}
\textcolor{mbzuai-c} {\Large\textbullet} mbzuai & 78.0 [58.6 - 86.4] & 0.67 [0.48 - 1.12] & 0.81 [0.58 - 1.45] \\
\\ \bottomrule
\end{tabular}
\end{table*}

\subsection{Evaluation per structure and impact on the rank}

The robustness and performance of the proposed techniques vary significantly depending on the structure, which directly influences the overall ranking. Evaluating the distribution of metrics is essential to understanding the stability and reliability of these methods. Notably, greater variability in algorithm performance is observed for the VS structure compared to the cochlea in both the 2022 and 2023 editions. As shown in \figref{fig:vs_dice} and \figref{fig:cochlea_dice}, more outliers are present for VS than for the cochlea according to the DSC metric. Similarly, \figref{fig:vs_assd} and \figref{fig:cochlea_assd} reveal a higher number of outliers for VS compared to the cochlea, as measured by the ASSD metric. This observation aligns with the 2021 edition observations, with proposed algorithms been less robust on VS than on cochleas. This can be explained by the uniformity of cochleas in terms of location, volume size, and intensity distribution, as compared to tumours.
\tabref{tab:cumulative_ranking} presents the distribution of cumulative ranks for each structure, VS and cochleas.
\begin{table*}[ht]
\centering
\caption{Distribution of individual cumulative ranks for VS and cochlea (median and IQR) in 2022 edition; and for intra-meatal VS, extra-meatal VS and cochlea (median and IQR) in 2023 edition.}
\label{tab:cumulative_ranking}
\begin{tabular}{lcccc}
\midrule
\textbf{Team} & \textbf{Challenge Rank} & \multicolumn{2}{c}{\textbf{VS}} & \textbf{Cochlea} \\
\hline
\textcolor{ne2e-c} {\Large\textbullet} ne2e & 1 & \multicolumn{2}{c}{3.5[1.5 - 6.0]} & 1.5[1.0 - 2.5] \\
\rowcolor{LightGray}
\textcolor{MAI-c} {\Large\textbullet} MAI & 2 & \multicolumn{2}{c}{4.0[2.0 - 5.5]} & 3.0[2.0 - 4.0] \\
\textcolor{LaTIM-c} {\Large\textbullet} LaTIM & 3 & \multicolumn{2}{c}{3.5[2.0 - 5.0]} & 4.0[2.5 - 5.5] \\
\rowcolor{LightGray}
\textcolor{Super-Polymerization-c} {\Large\textbullet} Super\textunderscore Polymerization & 4 & \multicolumn{2}{c}{4.0[2.5 - 5.5]} & 4.0[3.0 - 5.5] \\
\textcolor{A*DA-c} {\Large\textbullet} A*DA & 5 & \multicolumn{2}{c}{4.5[2.0 - 6.0]} & 5.5[4.5 - 6.5] \\
\rowcolor{LightGray}
\textcolor{fgh-365-c} {\Large\textbullet} fgh\textunderscore 365 & 6 & \multicolumn{2}{c}{7.0[5.5 - 9.0]} & 4.0[3.0 - 5.5] \\
\textcolor{SJTU-EIEE-2-426Lab-c} {\Large\textbullet} SJTU\textunderscore EIEE\textunderscore 2-426Lab & 7 & \multicolumn{2}{c}{5.5[3.0 - 7.5]} & 7.0[6.0 - 7.5] \\
\rowcolor{LightGray}
\textcolor{MBZUAI-VS-c} {\Large\textbullet} MBZUAI\textunderscore VS & 8 & \multicolumn{2}{c}{6.5[4.5 - 7.5]} & 9.5[9.0 - 10.0] \\
\textcolor{HUST-CBIB-c} {\Large\textbullet} HUST\textunderscore CBIB & 9 & \multicolumn{2}{c}{9.5[8.0 - 10.0]} & 8.5[8.0 - 9.5] \\
\rowcolor{LightGray}
\textcolor{skjp-c} {\Large\textbullet} skjp & 10 & \multicolumn{2}{c}{11.0[10.0 - 11.0]} & 8.5[7.5 - 9.5] \\
\textcolor{gabybaldeon-c} {\Large\textbullet} gabybaldeon & 11 & \multicolumn{2}{c}{9.0[8.5 - 10.0]} & 11.5[11.5 - 12.0] \\
\rowcolor{LightGray}
\textcolor{Of-Men-and-Rabbits-c} {\Large\textbullet} Of\textunderscore Men\textunderscore and\textunderscore Rabbits & 12 & \multicolumn{2}{c}{12.0[11.5 - 12.0]} & 11.0[11.0 - 11.5] \\ \midrule
\textbf{Team} & \textbf{Challenge Rank} & \textbf{Intra-meatal VS} & \textbf{Extra-meatal VS} & \textbf{Cochlea} \\ \midrule

\textcolor{vandy365-c} {\Large\textbullet} vandy365 & 1 & 1.5[1.0 - 2.5] & 1.0[1.0 - 2.0] & 2.5[1.5 - 3.5] \\
\rowcolor{LightGray}
\textcolor{ccc1018-c} {\Large\textbullet} ccc1018 & 2 & 2.5[2.0 - 4.0] & 3.0[2.0 - 4.0] & 2.5[1.5 - 4.0] \\
\textcolor{superpoly-c} {\Large\textbullet} superpoly & 3 & 4.0[3.0 - 5.0] & 3.0[1.5 - 4.0] & 2.0[1.0 - 3.5] \\
\rowcolor{LightGray}
\textcolor{hustcbib-c} {\Large\textbullet} hustcbib & 4 & 3.0[2.5 - 4.0] & 3.0[2.0 - 4.0] & 4.0[3.0 - 5.0] \\
\textcolor{auda-c} {\Large\textbullet} auda & 5 & 4.5[3.0 - 5.0] & 5.0[4.0 - 6.0] & 4.0[2.0 - 5.0] \\
\rowcolor{LightGray}
\textcolor{mbzuai-c} {\Large\textbullet} mbzuai & 6 & 5.0[4.0 - 6.0] & 5.0[4.0 - 6.0] & 5.5[5.0 - 6.0] \\
\hline
\end{tabular}
\end{table*}

\subsection{Remarks about the ranking stability}
Challenge rankings are influenced by several key design factors, including the choice of the test set, the performance metrics used, and the aggregation method applied \citep{maier2018rankings}. In the 1st edition of \crossmoda, we demonstrated the robustness and stability of our ranking methodology against these factors \citep{dorent2023crossmoda}. In this section, we reevaluate these aspects for the 2nd and 3rd editions to ensure consistency and reliability across different editions.

\begin{figure*}[htb!]
\begin{center}
\includegraphics[width=0.9\textwidth]{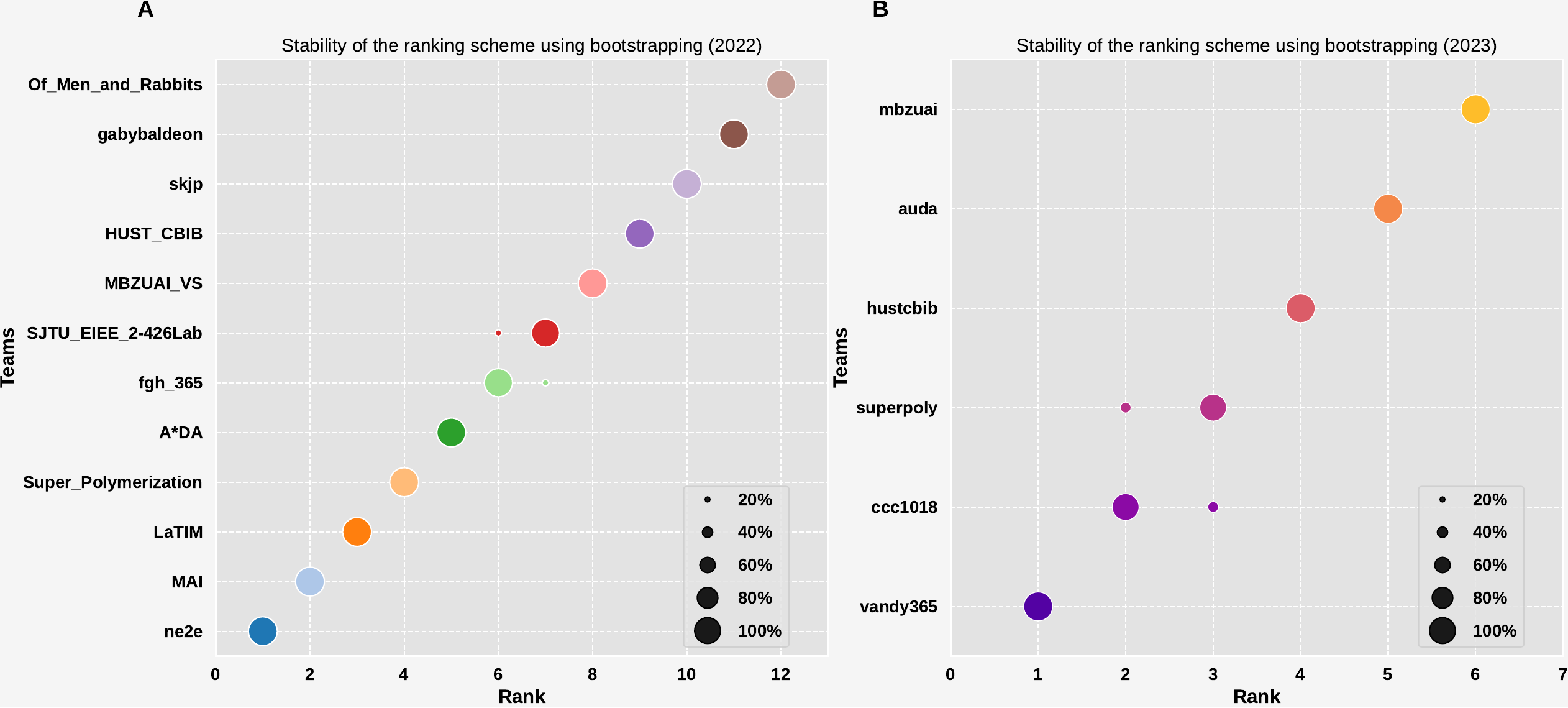}
\end{center}
\caption{Stability of the ranking scheme with respect to the choice of the metrics \textbf{(A)} 2022 \textbf{(B)} 2023 editions. 1000 bootstrap samples are used.}\label{fig:stability_blob}
\end{figure*}

Leveraging an established approach for assessing ranking stability \citep{wiesenfarth2021methods,dorent2023crossmoda}, we employed bootstrapping with 1000 samples to measure ranking variability. For each bootstrap sample, the ranking strategy was applied independently, and Kendall's $\tau$ was calculated to quantify agreement between the original ranking and the bootstrap-derived rankings. Kendall's $\tau$ ranges from -1 (complete disagreement) to 1 (perfect agreement).
The median Kendall's $\tau$ was 1 and the corresponding IQR was [1; 1], indicating perfect ranking stability.  
This is visualised in \figref{fig:stability_blob}, where the blob plot highlights the consistency of rankings across bootstrap samples. The top-ranked team maintained its position across all bootstrap samples for both \crossmoda 2022 and 2023 editions. 

\begin{figure*}[htb!]
\begin{center}
\includegraphics[width=0.8\textwidth]{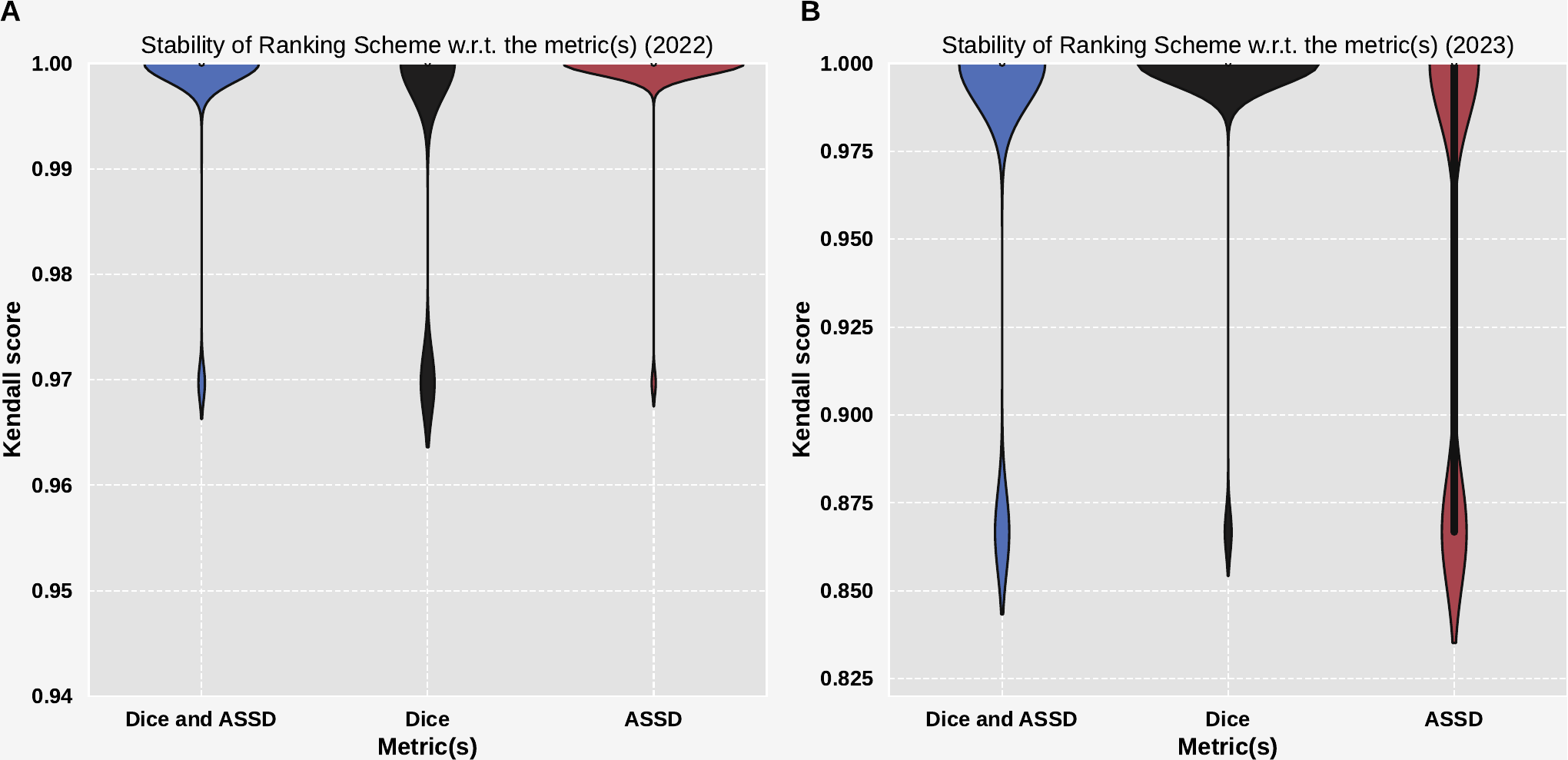}
\end{center}
\caption{Stability of the ranking scheme with respect to the choice of the metrics \textbf{(A)} 2022 \textbf{(B)} 2023 editions. 1000 bootstrap samples are used.}\label{fig:stability_metric}
\end{figure*}

To investigate the influence of metric choice on ranking stability, we compared single-metric approaches (DSC or ASSD) with a combined multi-metric strategy (DSC and ASSD). Using bootstrapping, Kendall's $\tau$ values were computed for each configuration. \figref{fig:stability_metric} shows a higher dispersion of Kendall's  $\tau$ values in single-metric approaches compared to the multi-metric approach. For \crossmoda 2022 and 2023 editions, Median Kendall's $\tau$ values were 1, 1, and 1 for DSC, ASSD, and the combined metric, respectively, indicating perfect ranking stability across the metrics evaluated.  %%% can explain the plot 

Additionally, we compared our ranking strategy with other aggregation methods. The two main approaches considered were:
\begin{itemize}
    \item Aggregate-then-rank: Metric values are aggregated (e.g., mean or median) across test cases, and then ranks are computed for each team.
    \item Rank-then-aggregate: Ranks are first determined per test case and metric, then aggregated (e.g., mean or median) to derive final rankings.
\end{itemize}

\begin{figure*}[htb!]
\begin{center}
\includegraphics[width=1.0\textwidth]{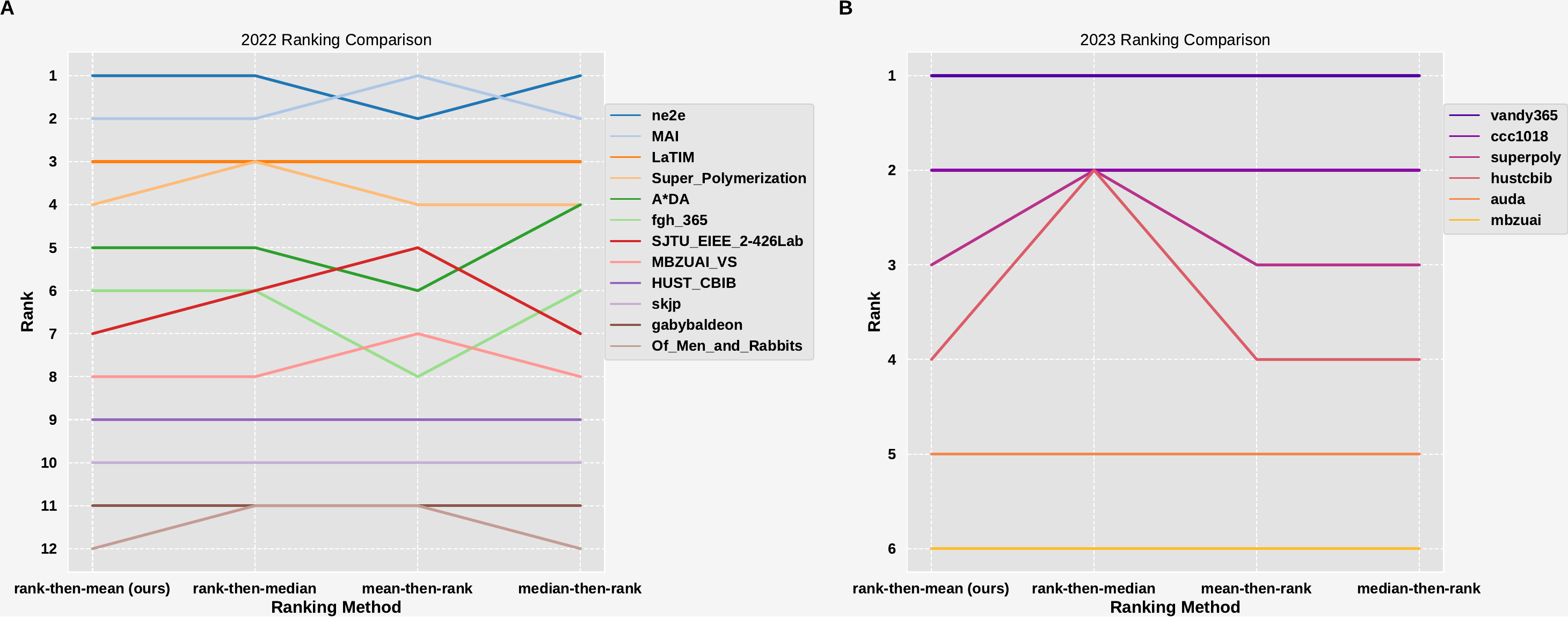}
\end{center}
\caption{Line plots visualising rankings robustness across different ranking methods for the brain task for \textbf{(A)} 2022 \textbf{(B)} 2023 editions. Each algorithm is represented by one coloured line. For each ranking method encoded on the x-axis, the height of the line represents the corresponding rank. The lowest rank of tied values is used for ties (equal scores for different teams).}\label{fig:ranking_comparison}
\end{figure*}

Our approach follows a rank-then-aggregate strategy using the mean for aggregation. Comparisons were made with a rank-then-aggregate method using the median and two aggregate-then-rank methods (mean and median aggregation).
\figref{fig:ranking_comparison} illustrates ranking consistency across these approaches. The aggregate-then-rank approach using the mean showed reduced robustness, particularly due to outliers in the ASSD metric. Overall, our findings demonstrate that the proposed ranking methodology is both stable and reliable, reinforcing confidence in the results.

\section{Discussion}

\subsection{Performance of automated segmentation methods from 2021 to 2023}
\paragraph{Synthesis approaches}
The winning teams from 2021, 2022, and 2023 editions of the competition placed a strong emphasis on enhancing the synthesis of synthetic T2 images to achieve high Dice scores. In 2021, the first-place team introduced segmentation decoders into the 2D CycleGAN~\citep{zhu2017unpaired} generators to better preserve intricate structures, particularly the small and often subtle cochlea. This modification enabled the model to more accurately capture and represent the structures of interest in the segmentation task. This approach was adopted and further refined by the winning teams in the 2022 and 2023 editions, with the synthesis network being extended to a 3D architecture. 

\paragraph{Augmentation techniques}
To further enhance the segmentation performance, the top teams also employed various augmentation techniques. 
In 2021, the third-place team used offline data augmentation to simulate heterogeneous signal intensities of the VS by altering the tumour intensities. This approach helped better prepare the model for the variability found in real-world data. 
Similarly, in 2022, the third place team used SinGAN~\citep{singan} for intensity augmentation, which generated realistic variations of VS appearances, making the model more robust to different tumour intensities. This approach in augmentation of local intensity proved effective in handling the variability inherent in the data and contributed to improved segmentation results. In 2023, this technique was employed by the winning team, in addition to introducing the oversampling of the hard samples. 

Additionally, image translation-based augmentation was introduced in the 2022 edition by the third-place team, who incorporated SinGAN blending to address multi-institutional data variability. This approach was further refined in the 2023 edition by the first and second-place teams, who introduced site-specific generative augmentation techniques to handle more heterogeneous data in the 2023 edition. 

\paragraph{Segmentation approaches}

The default 3D nnUNet network~\citep{isensee2021nnu}, including its preprocessing and augmentation techniques, was widely adopted by teams for the segmentation task in all editions. nnUNet \citep{isensee2021nnu} set the benchmark for segmentation, serving as the gold standard in VS and cochlea delineation on MRI.

\begin{table*}[htb]
\centering
\caption{Distribution of individual cumulative ranks for VS and cochlea (median and IQR); L. SC-GK: \londonscgk, T. SC-GK: \tilburgscgk}
\label{tab:comparison21-23}
\begin{tabular}{p{9cm}P{3cm}P{4cm}}
\toprule
\textbf{Winners' performance across editions} & \textbf{DSC (\%)} & \textbf{ASSD (mm)} \\
\midrule
\multicolumn{3}{l}{\textbf{VS}}\\ \midrule
\rowcolor{LightGray}2021 winner L. SC-GK test set &  87.01[81.81-89.99] &  0.3914[0.2869-0.5207]\\
2022 winner L. SC-GK test set & 88.45[84.71-91.95]  & 0.3041[0.2285-0.3798] \\
\rowcolor{LightGray}2023 winner L. SC-GK test set & 88.61[85.78-91.71] & 0.3065[0.2549-0.3854] \\
2022 winner L. SC-GK + T. SC-GK test set & 86.07[82.70-89.66]  & 0.3847[0.2794-0.6115] \\
\rowcolor{LightGray}2023 winner L. SC-GK + T. SC-GK test set & 88.00[84.27-90.90] & 0.3765[0.2921-0.4756] \\
2023 winner L. SC-GK + T. SC-GK + \londonmcrc test set & 87.08[83.25-90.22] & 0.4004[0.3162-0.5130]\\ \midrule
\multicolumn{3}{l}{\textbf{Cochlea}}\\ \midrule
\rowcolor{LightGray}2021 winner L. SC-GK test set & 85.71[83.85-86.98] & 0.1329[0.1174-0.1661]  \\
2022 winner L. SC-GK test set & 87.11[85.04-88.16] & 0.1249[0.1093-0.1537] \\
\rowcolor{LightGray}2023 winner L. SC-GK test set & 86.30[84.91-87.35] & 0.1177[0.1037-0.1330] \\
2022 winner L. SC-GK + T. SC-GK test set & 87.63[86.31-88.68] & 0.1469[0.1230-0.1701] \\
\rowcolor{LightGray}2023 winner L. SC-GK + T. SC-GK test set & 84.90[82.97-86.53] & 0.1775[0.1207-0.2350] \\
2023 winner L. SC-GK + T. SC-GK + \londonmcrc test set & 84.06[81.65-85.95] & 0.2074[0.1330-0.2467] \\ \bottomrule
\end{tabular}
\end{table*}

\subsection{Retrospective analysis of the top-ranked methods from 2021 to 2023}
In order to fairly assess the performance of winning models of the each edition across datasets used in each editions, we perform a retrospective analysis taking into account the variations in datasets and evolving task objectives. 
\begin{figure*}[htb!]
\begin{center}
\includegraphics[width=0.9\textwidth]{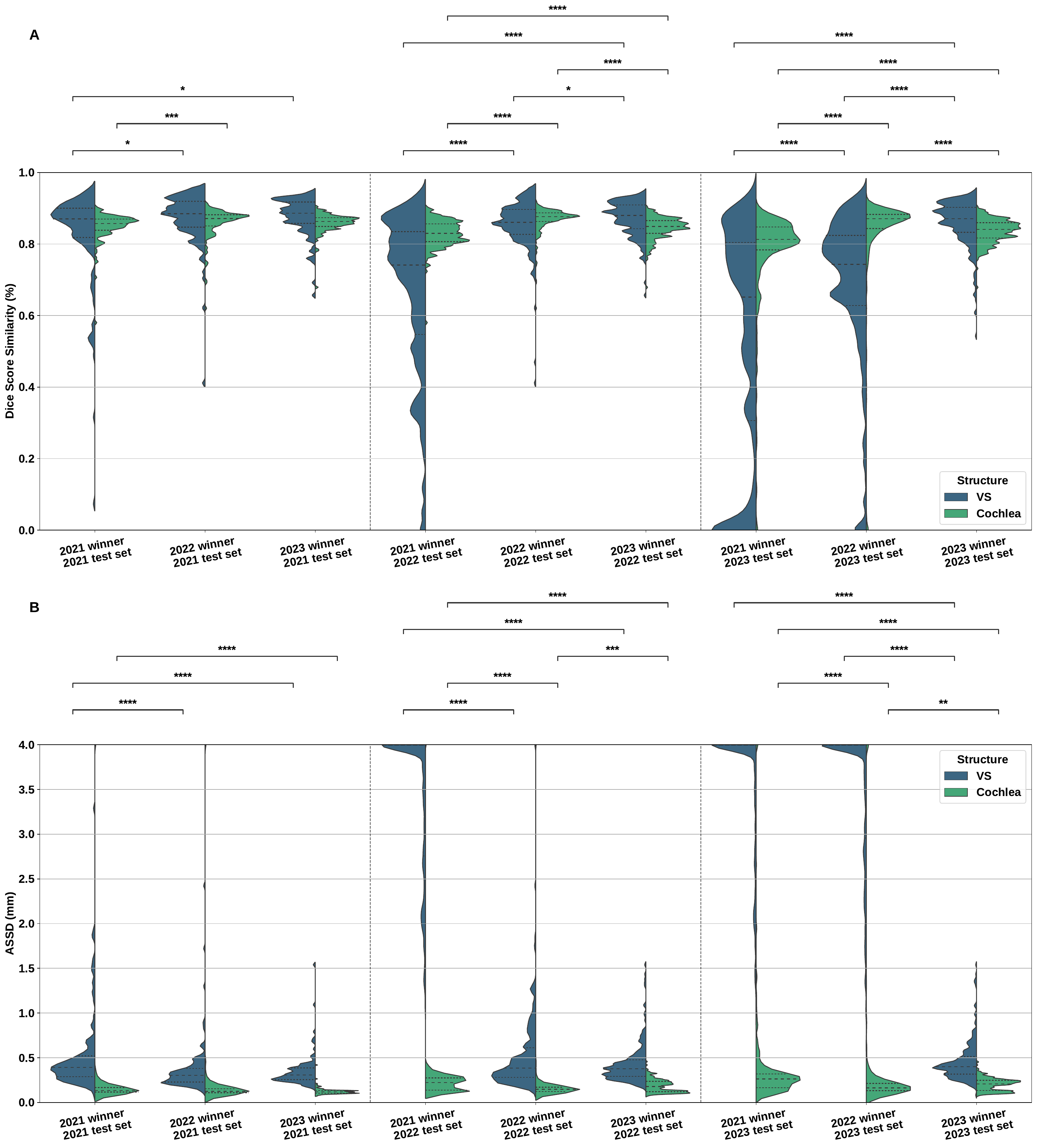}
\end{center}
\caption{Comparison of the overall performance by the latest winning teams across previous editions of the \crossmoda challenge based on (A) Dice score (B) ASSD metric for the two structures VS and cochlea}\label{fig:comparison21-23}
\end{figure*}

\figref{fig:comparison21-23} illustrates the performance across the \crossmoda challenge editions from 2021 to 2023, focusing on the key evaluation metrics: Dice score and ASSD. \figref{fig:comparison21-23}(A) showcases the Dice score for the VS and cochlea segmentation across various datasets used in each edition, highlighting a steady performance improvement for VS, especially notable with the 2023 winner model trained with dataset incorporating the most diverse, multi-institutional data. 2023 winning model trained on most diverse data, show a significant improvement in VS segmentation on the \londonscgk compared to the 2021 winner model trained only on \londonscgk data ($p<0.01$). 2022 winning model trained on two single institute datasets, show a significant improvement in VS segmentation  on the \londonscgk compared to the 2021 winner model trained only on \londonscgk ($p<0.01$). Less outliers are observed within the performance on the single institute \londonscgk dataset for the 2023 winner model compared to 2022 and 2021 winning models. 2023 winner shows a significant improvement in Dice scores for the VS structure on the \londonscgk + \tilburgscgk combined, compared to the 2022 winner trained on the same \londonscgk + \tilburgscgk datasets. Overall the VS performance is more than 88\% on single institute datasets for 2023 winner model, even though the diverse test set performance is 87.08\%. However a reduced outlier spread and improved median performance metrics can be observed for the VS structure segmentation with regards to Dice score over time for the 2023 winning model. 

As seen in \figref{fig:comparison21-23}(A), the 2021 winning model demonstrates a notable performance drop when evaluated on the 2022 and 2023 datasets.
The substantial number of outliers indicates poor generalisation to data outside its original training distribution. 
Similarly, the 2022 winner model, which was trained on two single-institute datasets, shows reduced performance on the 2023 dataset. This decline emphasises a key observation, models trained on homogeneous, single-source data struggle to generalise effectively to heterogeneous, multi-institutional test sets. In contrast, the 2023 winner model, trained on the most diverse and multi-institutional dataset, maintains high performance across all test sets, including the more varied 2023 data, with fewer outliers and more consistent Dice scores. These results highlight the importance of diverse training data in developing robust segmentation models capable of generalising well across different institutions and datasets.

For cochlea segmentation, a Dice score improvement is observed from the 2021 to the 2022 winner models on the \londonscgk dataset, with the 2022 winner achieving significantly higher scores compared to the 2021 winner ($p<0.005$). However, the 2023 winner model shows a slight decrease in median Dice scores for the cochlea on the \londonscgk dataset compared to the 2022 winner, likely due to the increased complexity introduced by multi-institutional training data and increased challenge posed by the new intra/extra-VS sub-segmentation requirements. Similarly, on the combined \londonscgk + \tilburgscgk datasets, the 2023 winner exhibits a significant drop in Dice scores ($p<0.005$) compared to the 2022 winner trained on the same two datasets, further suggesting that the increased data diversity challenges precision for small anatomical structures like the cochlea.  

\figref{fig:comparison21-23} (B) presents ASSD values, indicating a significant reduction in surface distance errors for both structures over the years, underscoring the efficacy of models trained on heterogeneous data. The figure overall emphasises that incorporating diverse datasets not only improves segmentation generalisability but also refines predictions on homogeneous and challenging cases alike. 

The decrease in cochlea segmentation performance in the 2023 edition warrants particular attention. Unlike the tumour labels, the cochlea annotation target did not change conceptually across editions, making this trend less straightforward to interpret. The observed decline should therefore not be attributed solely to the introduction of intra-/extra-meatal tumour sub-segmentation. Several factors may have contributed. First, the 2023 edition incorporated the \londonmcrc dataset, which introduced greater variability in voxel volume, slice thickness, intensity distribution, scanner vendor, and acquisition protocol. Such variability may affect small anatomical structures such as the cochlea, for which small boundary errors can produce large DSC changes. Second, the transition from a two-class to a three-class segmentation task may have altered the optimisation balance between structures, particularly because the intra-meatal tumour component and cochlea are spatially close and both relatively small. Third, methods optimised primarily for tumour robustness under heterogeneous routine imaging may have traded off some precision for cochlear delineation.

Nonetheless, the present analysis does not fully resolve the mechanism underlying the cochlea performance decline. Further work should stratify performance by dataset source, imaging characteristics, and tumour location, and investigate interactions between tumour sub-segmentation and cochlea predictions. We therefore consider this an open question for future analyses.

Another limitation is the absence of demographic metadata in the released challenge data. Although the datasets capture substantial imaging and institutional heterogeneity, the lack of age, sex, and ethnicity information prevents analysis of demographic subgroup performance. Future challenge datasets should consider including harmonised and appropriately de-identified demographic variables to support more comprehensive fairness and generalisability analyses.

Several barriers remain before cross-modality DA methods can be adopted in clinical practice, including the need for consistent performance across heterogeneous data, reduction of outliers, validation of clinical utility beyond segmentation accuracy, and robust quality-control mechanisms. Prospective validation across institutions and integration into clinical workflows, alongside regulatory considerations, are also essential. Future challenges could address these aspects by incorporating clinically relevant endpoints, uncertainty estimation, more comprehensive metadata, and broader cross-modality settings such as MRI-to-intra-operative ultrasound adaptation to better assess generalisability and robustness.

\section{Conclusion}
In conclusion, the evolution of the \crossmoda challenge from 2021 to 2023 demonstrates progress in addressing cross-modality segmentation for VS and cochlea. Each edition expanded its scope, moving from single-institutional, homogeneous data to multi-institutional, heterogeneous datasets, and adding tasks such as grading VS and distinguishing tumour components. These changes have advanced the clinical relevance of the challenge and provided insights into the impact of data diversity and task complexity on segmentation performance. The reduction in outliers with increased data heterogeneity shows the potential for improved segmentation accuracy on homogeneous datasets. However, the decrease in cochlea Dice scores in the 2023 edition highlights the challenges of maintaining performance across all segmentation classes. Overall, the \crossmoda challenge continues to drive progress in medical image analysis, providing a platform for developing techniques that address domain shifts in clinical imaging. Looking ahead, we aim to broaden the challenge to other imaging modalities beyond MRI, with a particular focus on ultrasound. Leveraging the ReMIND dataset~\citep{juvekar2024remind}, we for example intend to investigate cross-modality adaptation between MRI and intra-operative ultrasound, further increasing the domain gap across domains. This shift will enable the evaluation of algorithms under more severe distributional shifts and support the development of models for image-guided interventions.

\section*{Acknowledgments}
N. Wijethilake was supported by the UK Medical Research Council [MR/N013700/1] and the King’s College London MRC Doctoral Training Partnership in Biomedical Sciences. This work was supported by core funding from the Wellcome Trust (203148/Z/16/Z) and EPSRC (NS/A000049/1) and an MRC project grant (MC/PC/180520). 
R. Dorent received a Marie
Skłodowska-Curie grant No 101154248 (project: SafeREG).  TV is also supported by a Medtronic/Royal Academy of Engineering Research Chair (RCSRF1819/7/34). For the purpose of open access, the authors have applied a CC BY public copyright licence to any Author Accepted Manuscript version arising from this submission.

\section*{Contributions} 

\textbf{Navodini Wijethilake:} Conceptualization, Methodology, Software, Formal analysis, Resources, Data curation, Writing – original draft, Writing – review \& editing, Visualization. 
\textbf{Reuben Dorent:} Conceptualization, Methodology, Software, Formal analysis, Resources, Data curation, Writing – original draft, Writing – review \& editing, Visualization. 
\textbf{Marina Ivory:} Conceptualization, Methodology, Data curation. 
\textbf{Aaron Kujawa:} Conceptualization, Methodology. 
\textbf{Stefan Cornelissen:} Methodology, Data curation. 
\textbf{Patrick Langenhuizen:} Methodology, Data curation. 
\textbf{Mohamed Okasha:} Methodology, Data curation. 
\textbf{Anna Oviedova:} Methodology, Data curation. 
\textbf{Hexin Dong:} Methodology, Software. 
\textbf{Bogyeong Kang:} Methodology, Software. 
\textbf{Guillaume Sall\'e:} Methodology, Software. 
\textbf{Luyi Han:} Methodology, Software. 
\textbf{Ziyuan Zhao:} Methodology, Software. 
\textbf{Han Liu:} Methodology, Software. 
\textbf{Yubo Fan:} Methodology, Software. 
\textbf{Tao Yang:} Methodology, Software. 
\textbf{Shahad Hardan:} Methodology, Software. 
\textbf{Hussain Alasmawi:} Methodology, Software. 
\textbf{Santosh Sanjeev:} Methodology, Software. 
\textbf{Yuzhou Zhuang:} Methodology, Software. 
\textbf{Satoshi Kondo:} Methodology, Software. 
\textbf{Maria Baldeon Calisto:} Methodology, Software. 
\textbf{Shaikh Muhammad Uzair Noman:} Methodology, Software. 
\textbf{Cancan Chen:} Methodology, Software. 
\textbf{Ipek Oguz:} Methodology, Software. 
\textbf{Rongguo Zhang:} Methodology, Software. 
\textbf{Mina Rezaei:} Methodology, Software. 
\textbf{Susana K. Lai-Yuen:} Methodology, Software. 
\textbf{Satoshi Kasai:} Methodology, Software. 
\textbf{Chih-Cheng Hung:} Methodology, Software. 
\textbf{Mohammad Yaqub:} Methodology, Software. 
\textbf{Lisheng Wang:} Methodology, Software. 
\textbf{Benoit M. Dawant:} Methodology, Software. 
\textbf{Cuntai Guan:} Methodology, Software. 
\textbf{Ritse Mann:} Methodology, Software. 
\textbf{Vincent Jaouen:} Methodology, Software. 
\textbf{Ji-Wung Han:} Methodology, Software. 
\textbf{Li Zhang:} Methodology, Software. 
\textbf{Jonathan Shapey:} Conceptualization, Methodology, Data curation, Writing – review \& editing, Funding acquisition. 
\textbf{Tom Vercauteren:} Project administration, Conceptualization, Methodology, Formal analysis, Resources, Writing – original draft, Writing – review \& editing, Funding acquisition.
%\FloatBarrier

%\section*{References}
% \bibliography{refs}
% \section*{\itshape Reference style}

% Text: All citations in the text should refer to:
% \begin{enumerate}
% \item Single author: the author's name (without initials, unless there
% is ambiguity) and the year of publication;
% \item Two authors: both authors' names and the year of publication;
% \item Three or more authors: first author's name followed by `et al.'
% and the year of publication.
% \end{enumerate}
% Citations may be made directly (or parenthetically). Groups of
% references should be listed first alphabetically, then chronologically.

% %%Harvard
\bibliographystyle{model2-names.bst}\biboptions{authoryear}
\bibliography{refs}

% \section*{Supplementary Material}

% Supplementary material that may be helpful in the review process should
% be prepared and provided as a separate electronic file. That file can
% then be transformed into PDF format and submitted along with the
% manuscript and graphic files to the appropriate editorial office.

\end{document}